\pdfoutput=1
\newcommand*{\ATLASLATEXPATH}{latex/}

\documentclass[UKenglish,texlive=2016,cernpreprint,paper=a4]{\ATLASLATEXPATH atlasdoc}

\usepackage[backend=biber]{\ATLASLATEXPATH atlaspackage}
\usepackage{\ATLASLATEXPATH atlasbiblatex}

\usepackage{\ATLASLATEXPATH atlascontribute}

\usepackage{\ATLASLATEXPATH atlasphysics}

\graphicspath{{logos/}{figures/}}

\usepackage{HIGG-2016-15-PAPER-defs}
\usepackage{dcolumn}
\usepackage{multirow}
\usepackage{siunitx}
\AtlasTitle{Search for Higgs boson pair production in the \yybb final state with \SI{13}{\TeV} \pp collision data collected by the ATLAS experiment}

\author{The ATLAS Collaboration}

\AtlasRefCode{HIGG-2016-15}

\PreprintIdNumber{CERN-EP-2018-130}

\AtlasDate{13 Jul 2018}

\arXivId{1807.04873}

\HepDataRecord{82818}

\AtlasJournal{JHEP}
\AtlasJournalRef{JHEP 11 (2018) 040}
\AtlasDOI{10.1007/JHEP11(2018)040}

\AtlasAbstract{%
A search is performed for resonant and non-resonant Higgs boson pair production in the \yybb final state.
The data set used corresponds to an integrated luminosity of \lumiInFemtoBarn of proton--proton collisions at a centre-of-mass energy of \SI{13}{\TeV} recorded by the ATLAS detector at the CERN Large Hadron Collider.
No significant excess relative to the Standard Model expectation is observed.
The observed limit on the non-resonant Higgs boson pair \xs is \SI{\nonresobs}{\pico\barn} at 95\% confidence level.
This observed limit is equivalent to \num{\nonresnSigmaobs} times the predicted Standard Model \xs.
The Higgs boson self-coupling ($\khhh = \lhhh / \lhhhSM$) is constrained at 95\% confidence level to $\lambdaminobs < \khhh < \lambdamaxobs$.
For resonant Higgs boson pair production through \Xhhyybb, the limit is presented, using the narrow-width approximation, as a function of \mX in the range $\SI{260}{\GeV} < \mX < \SI{1000}{\GeV}$.
The observed limits range from \SI{\resmaxobs}{\pico\barn} to \SI{\resminobs}{\pico\barn} over this mass range.
}

\hypersetup{pdftitle={ATLAS paper HIGG-2016-15},pdfauthor={The ATLAS Collaboration}}

\begin{document}

\maketitle

\section{Introduction}\label{sec:introduction}
\begin{figure}[!htb]
    \centering
    \subfloat[]{\includegraphics[height=0.1\textheight]{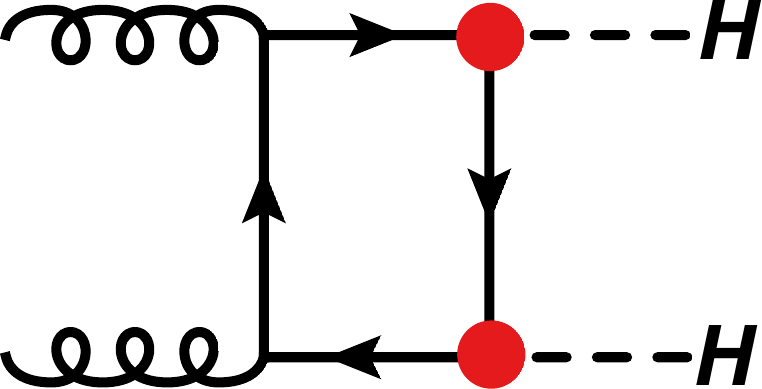}}
    \quad
    \subfloat[]{\includegraphics[height=0.1\textheight]{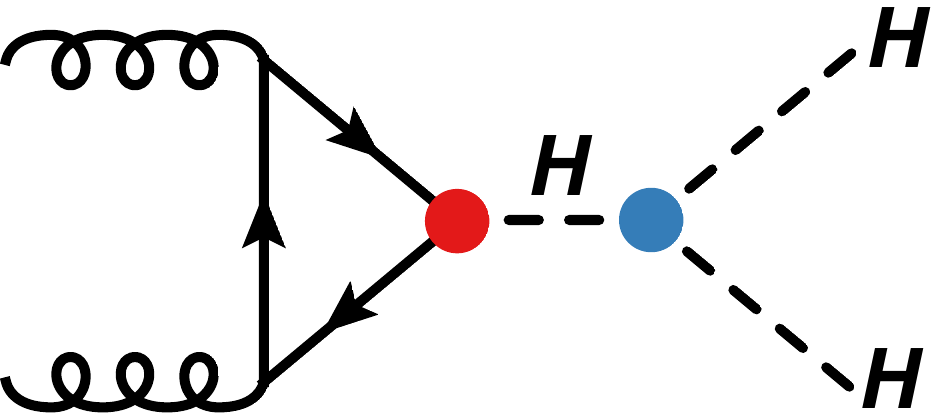}}
    \quad
    \subfloat[]{\includegraphics[height=0.1\textheight]{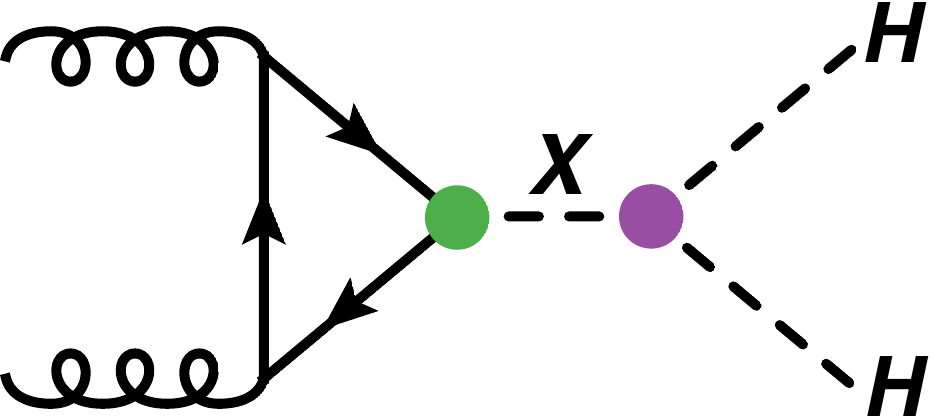}}
    \caption{Leading-order production modes for Higgs boson pairs.
        In the SM, there is destructive interference between (a) the heavy-quark loop and (b) the Higgs self-coupling production modes, which reduces the overall \xs.
        BSM Higgs boson pair production could proceed through changes to the Higgs couplings, for example the \ttH or $HHH$ couplings which contribute to (a) and (b), or through an intermediate resonance, $X$, which could, for example, be produced through a quark loop as shown in (c).}
    \label{fig:intro-Higgs-pair-production}
\end{figure}

The Higgs boson ($H$) was discovered by the ATLAS~\cite{HIGG-2012-27} and CMS~\cite{CMS-HIG-12-028} collaborations in 2012 using proton--proton (\pp) collisions at the Large Hadron Collider (LHC).
Measurements of the properties of the boson are in agreement with the predictions of the Standard Model (SM)~\cite{HIGG-2014-14,HIGG-2015-07}.
If SM expectations hold, the production of a Higgs boson pair in a single \pp interaction should not be observable with the currently available LHC data set.
In the SM, the dominant contributions to this process are shown in \Figs{\ref{fig:intro-Higgs-pair-production}(a)}{\ref{fig:intro-Higgs-pair-production}(b)}.
However, some beyond-the-Standard-Model (BSM) scenarios may enhance the Higgs boson pair production rate.

Many BSM theories predict the existence of heavy particles that can decay into a pair of Higgs bosons.
These could be identified as a resonance in the Higgs boson pair invariant mass spectrum.
They could be produced, for example, through the gluon--gluon fusion mode shown in \Fig{\ref{fig:intro-Higgs-pair-production}(c)}.
Models with two Higgs doublets~\cite{Lee:1973iz}, such as the minimal supersymmetric extension of the SM~\cite{Dimopoulos:1981zb}, twin Higgs models~\cite{Chacko:2005vw} and composite Higgs models~\cite{Grober:2010yv,Mrazek:2011iu}, add a second complex scalar doublet to the Higgs sector.
In general, the neutral Higgs fields from the two doublets will mix, which may result in the existence of a heavy Higgs boson that decays into two of its lighter Higgs boson partners.
Alternatively, the Randall--Sundrum model of warped extra dimensions~\cite{Randall:1999ee} predicts spin-0 radions and spin-2 gravitons that could couple to a Higgs boson pair.

In addition to the resonant production, there can also be non-resonant enhancements to the Higgs boson pair \xs.
These can either originate from loop corrections involving new particles, such as light, coloured scalars~\cite{Kribs:2012kz}, or through non-SM couplings.
Changes to the single Higgs boson production \xs arising from such loop-corrections are neglected in this paper.
Anomalous couplings can either be extensions to the SM, such as contact interactions between two top quarks and two Higgs bosons~\cite{Contino:2012xk}, or be deviations from the SM values of the couplings between the Higgs boson and other particles.
In this work, the effective Higgs self-coupling, \lhhh, is parameterised by a scale factor \khhh ($\khhh = \lhhh / \lhhhSM$) where the SM superscript refers to the SM value of this parameter.
The theoretical and phenomenological implications of such couplings for complete models are discussed in \Refs{\cite{DiLuzio:2017tfn}}{\cite{Kribs:2017znd}}.
The Yukawa coupling between the top quark and the Higgs boson is set to its SM value in this paper, consistent with its recent direct observation~\cite{CMS-HIG-17-001,HIGG-2018-13}.

This paper describes a search for the production of pairs of Higgs bosons in \pp collisions at the LHC.
The search is carried out in the \yybb final state, and considers both resonant and non-resonant contributions.
Previous searches were carried out by the ATLAS and CMS collaborations in the \yybb channel at $\sqs = \SI{8}{\TeV}$~\cite{HIGG-2013-29,CMS-HIG-13-032}, as well as in other final states~\cite{EXOT-2016-31,HIGG-2013-33, CMS-HIG-17-002,CMS-HIG-17-006} at both $\sqs = \SI{8}{\TeV}$ and $\sqs = \SI{13}{\TeV}$.

Events are required to have two isolated photons, accompanied by two jets with dijet invariant mass (\mjj) compatible with the mass of the Higgs boson, $\mH = \SI{125.09}{\GeV}$~\cite{HIGG-2014-14}.
At least one of these jets must be tagged as containing a $b$-hadron; events are separated into signal categories depending on whether one or both jets are tagged in this way.

\textit{Loose} and \textit{tight} kinematic selections are defined, where the tight selection is a strict subset of the loose one.
The searches for low-mass resonances and for non-SM values of the Higgs boson self-coupling both use the loose selection, as the average transverse momentum (\pT) of the Higgs bosons is lower in these cases~\cite{Kling:2016lay}.
The tight selection is used for signals where the Higgs bosons typically have larger average \pT, namely in the search for higher-mass resonances and in the measurement of SM non-resonant \hh production.

In the search for non-resonant production, the signal is extracted using a fit to the diphoton invariant mass (\myy) distribution of the selected events.
The signal consists of a narrow peak around \mH superimposed on a smoothly falling background.
Only the predominant gluon--gluon fusion production mode, which represents over 90\% of the SM \xs, is considered.

For resonant production, the signal is extracted from the four-object invariant mass (\myyjj) spectrum for events with a diphoton mass compatible with the mass of the Higgs boson, by fitting a peak superimposed on a smoothly changing background.
The narrow-width approximation is used, focusing on a resonance with mass (\mX) in the range $\SI{260}{\GeV} < \mX < \SI{1000}{\GeV}$.
Although this search is for a generic scalar decaying into a pair of Higgs bosons, the simulated samples used to optimise the search were produced in the gluon--gluon fusion mode.

The rest of this paper is organised as follows.
\Sect{\ref{sec:detector}} provides a brief description of the ATLAS detector, while \Sect{\ref{sec:data_MC}} describes the data and simulated event samples used.
An overview of object and event selection is given in \Sect{\ref{sec:selection}}, while \Sect{\ref{sec:modelling}} explains the modelling of signal and background processes.
The sources of systematic uncertainties are detailed in \Sect{\ref{sec:systematics}}.
Final results including expected and observed limits are presented in \Sect{\ref{sec:results}}, and \Sect{\ref{sec:conclusions}} summarises the main findings.


\newcommand{\AtlasCoordFootnote}{%
    ATLAS uses a right-handed coordinate system with its origin at the nominal interaction point (IP) in the centre of the detector and the $z$-axis along the beam pipe.
    The $x$-axis points from the IP to the centre of the LHC ring, and the $y$-axis points upwards.
    Cylindrical coordinates $(r,\phi)$ are used in the transverse plane, $\phi$ being the azimuthal angle around the $z$-axis.
    The pseudorapidity is defined in terms of the polar angle $\theta$ as $\eta = -\ln \tan(\theta/2)$.
    Angular distance is measured in units of $\Delta R \equiv \sqrt{(\Delta\eta)^{2} + (\Delta\phi)^{2}}$.}

\section{ATLAS detector}\label{sec:detector}
The ATLAS detector~\cite{PERF-2007-01} at the LHC is a multipurpose particle detector with a forward--backward symmetric cylindrical geometry\footnote{\AtlasCoordFootnote} and a near $4\pi$ coverage in solid angle.
It consists of an inner tracking detector (ID) surrounded by a thin superconducting solenoid providing a \SI{2}{\tesla} axial magnetic field, electromagnetic (EM) and hadronic calorimeters, and a muon spectrometer (MS).
The inner tracking detector, consisting of silicon pixel, silicon microstrip, and transition radiation tracking systems, covers the pseudorapidity range $|\eta| < 2.5$.
The innermost pixel layer, the insertable B-layer (IBL)~\cite{ATLAS-TDR-19}, was added between the first and second runs of the LHC, around a new, narrower and thinner beam pipe.
The IBL improves the experiment's ability to identify displaced vertices and thereby improves the performance of the \btagging algorithms~\cite{ATL-PHYS-PUB-2015-022}.
Lead/liquid-argon (LAr) sampling calorimeters with high granularity provide energy measurements of EM showers.
A hadronic steel/scintillator-tile calorimeter covers the central pseudorapidity range ($|\eta| < 1.7$), while a LAr hadronic endcap calorimeter provides coverage over $1.5<|\eta|<3.2$.
The endcap and forward regions are instrumented with LAr calorimeters for both the EM and hadronic energy measurements up to $|\eta| = 4.9$.
The MS surrounds the calorimeters and is based on three large air-core toroidal superconducting magnets, each with eight coils, and with bending power in the range of \num{2.0} to \SI{7.5}{\tesla\metre}.
It includes a system of precision tracking chambers, covering the region $|\eta| < 2.7$, and fast detectors for triggering purposes, covering the range $|\eta| < 2.4$.

A two-level trigger system is used to select interesting events~\cite{TRIG-2016-01}.
The first-level trigger is implemented in hardware and uses a subset of the total available information to make fast decisions to accept or reject an event, aiming to reduce the rate to around \SI{100}{\kilo\hertz}.
This is followed by the software-based high-level trigger (HLT), which runs reconstruction and calibration software, reducing the event rate to about \SI{1}{\kilo\hertz}.


\section{Data and simulated samples}\label{sec:data_MC}
\subsection{Data selection}
This analysis uses the \pp data sample collected at $\sqs = \SI{13}{\TeV}$ with the ATLAS detector in 2015 and 2016, corresponding to an integrated luminosity of \lumiInFemtoBarn.
All events for which the detector and trigger system satisfy a set of data-quality criteria are considered.
Events are selected using a diphoton trigger, which requires two photon candidates with transverse energy (\ET) above \num{35} and \SI{25}{\GeV}, respectively.
The overall trigger selection efficiency is greater than 99\% for events having the characteristics to satisfy the event selection detailed in \Sect{\ref{sec:selection}}.

\subsection{Simulated event samples}
\label{subsec:data_MC_samples}
Non-resonant production of Higgs boson pairs via the gluon--gluon fusion process was simulated at next-to-leading-order (\NLO) accuracy in QCD using an effective field theory (EFT) approach, with form factors for the top-quark loop from HPAIR~\cite{Dawson:1998py,Plehn:1996wb} to approximate finite top-quark mass effects.
The simulated events were reweighted to reproduce the \mhh spectrum obtained in \Refs{\cite{Borowka:2016ehy}}{\cite{Borowka:2016ypz}}, which calculated the process at \NLO in QCD while fully accounting for the top-quark mass.
The total \xs is normalised to \diHiggsXS, in accordance with a calculation at next-to-next-to-leading order (NNLO) in QCD~\cite{deFlorian:2016spz,Grazzini:2018bsd}.

Non-resonant BSM Higgs boson pair production with varied \khhh was simulated at \LO accuracy in QCD~\cite{Hespel:2014sla} for eleven values of \khhh in the range $-10 < \khhh < 10$.
The total cross-sections for these samples were computed as a function of \khhh at \LO accuracy in QCD.
A constant NNLO/LO $K$-factor (2.283) computed at $\khhh = 1$, was then applied.
As the amplitude for Higgs boson pair production can be expressed in terms of \khhh and the top quark's Yukawa coupling, weighted combinations of the simulated samples can produce predictions for other values of \khhh.

Resonant BSM Higgs boson pair production via a massive scalar, was simulated at \NLO accuracy for ten different mass points (\num{260}, \num{275}, \num{300}, \num{325}, \num{350}, \num{400}, \num{450}, \num{500}, \num{750} and \SI{1000}{\GeV}) using the narrow-width approximation.
For all generated Higgs boson pair samples, both resonant and non-resonant, the branching fractions for \hbb and \hyy and are taken to be 0.5809 and 0.00227 respectively~\cite{deFlorian:2016spz}.

This analysis is affected both by backgrounds from single-Higgs-boson production and by non-resonant backgrounds with continuum \myy spectra.
Background estimation is carried out using data-driven methods whenever possible; in particular, data are used to estimate the continuum background contribution from SM processes with multiple photons and jets, which constitute the dominant background for this search.
Monte Carlo event generators were used for the simulation of different signal hypotheses and the background from SM single Higgs boson production.
The major single Higgs boson production channels contributing to the background are gluon--gluon fusion (\ggH), associated production with a $Z$ boson (\ZH), associated production with a top quark pair (\ttH) and associated production with a single top quark (\tH).
In addition, contributions from vector-boson fusion (VBF H), associated production with a $W$ boson (\WH) and associated production with a bottom quark pair (\bbH) are also considered.
Overall, the largest contributions come from \ttH and \ZH.
More information about these simulated background samples can be found in \Ref{\cite{HIGG-2016-21}} and in \Tab{\ref{tab:data_MC-summary}}.

For all matrix element generators other than \SHERPA, the resulting events were passed to another program for simulation of parton showering, hadronisation and the underlying event.
This is either \HERWIGpp with the CTEQ6L1 parton distribution function (PDF) set~\cite{Pumplin:2002vw} using the UEEE5 set of tuned parameters~\cite{Gieseke:2012ft} or \PYTHIAV{8} with the NNPDF 2.3 LO PDF set~\cite{Ball:2012cx} and the A14 set of tuned parameters~\cite{ATL-PHYS-PUB-2014-021}.
For all simulated samples except those generated by \SHERPA, the \EVTGEN v1.2.0 program~\cite{Lange:2001uf} was used for modelling the properties of $b$- and $c$-hadron decays.
Multiple overlaid \pp collisions (pile-up) were simulated with the soft QCD processes of \PYTHIAV{8.186} using the A2 set of tuned parameters~\cite{ATL-PHYS-PUB-2012-003} and the MSTW2008LO PDF set~\cite{Martin:2009iq}.
The distribution of the number of overlaid collisions simulated in each event approximately matches what was observed during 2015 and 2016 data-taking.
Event-level weights were applied to the simulated samples in order to improve the level of agreement.

The final-state particles were passed either through a \GEANT~4~\cite{Agostinelli:2002hh} simulation of the ATLAS detector, or through the ATLAS fast simulation framework~\cite{SOFT-2010-01}, which has been extensively cross-checked against the \GEANT~4 model.
The output from this detector simulation step is then reconstructed using the same software as used for the data.
A list of the signal and dominant background samples used in the paper is shown in \Tab{\ref{tab:data_MC-summary}}.

\begin{table}[!htb]
    \begin{center}
        \caption{Summary of the event generators and PDF sets used to model the signal and the main background processes.
            The SM \xs{s} $\sigma$ for the Higgs boson production processes with \mH = \SI{125.09}{\GeV} are also given separately for $\sqs = \SI{13}{\TeV}$, together with the orders of the calculation corresponding to the quoted \xs{s}, which are used to normalise samples.
            The following generator versions were used: \PYTHIAV{8.212}~\cite{Sjostrand:2014zea} (event generation), \PYTHIAV{8.186}~\cite{Sjostrand:2007gs} (pile-up overlay); \HERWIGpp 2.7.1~\cite{Bahr:2008pv,Bellm:2013hwb}; \POWHEGBOX r3154 (base) v2~\cite{Nason:2004rx,Frixione:2007vw,Alioli:2010xd}; \MGMCatNLO 2.4.3~\cite{Alwall:2014hca}; \SHERPA 2.2.1~\cite{Gleisberg:2008ta,Hoeche:2009rj,Gleisberg:2008fv,Schumann:2007mg}.
            The PDF sets used are: CT10 NLO~\cite{Lai:2010vv}, CTEQ6L1~\cite{Pumplin:2002vw}, NNPDF 2.3 LO~\cite{Ball:2012cx}, NNPDF 3.0 LO~\cite{Ball:2014uwa}, PDF4LHC15~\cite{Butterworth:2015oua}.
            For the BSM signals, no cross-section is specified as it is the parameter of interest for measurement.
            For the \SHERPA background, no cross-section is used, as the continuum background is fit in data.
        }
        \label{tab:data_MC-summary}

        \resizebox{\textwidth}{!}
        {\normalsize
            \begin{tabular}{l l l l r l l}
                \toprule
                Process                    & Generator                                         & Showering   & PDF set      & $\sigma$ [fb]     & Order of calculation of $\sigma$ & Simulation \\
                \midrule
                Non-resonant SM \hh        & \MGMCatNLO                                        & \HERWIGpp   & CT10 NLO     & \diHiggsXSnoUnits & NNLO+NNLL                        & Fast       \\
                Non-resonant BSM \hh       & \MGMCatNLO                                        & \PYTHIAV{8} & NNPDF 2.3 LO & -                 & LO                               & Fast       \\
                Resonant BSM \hh           & \MGMCatNLO                                        & \HERWIGpp   & CT10 NLO     & -                 & NLO                              & Fast       \\
                \midrule
                \yy plus jets              & \SHERPA                                           & \SHERPA     & CT10 NLO     & -                 & LO                               & Fast       \\
                \midrule
                \ggH                       & \POWHEGBOX NNLOPS (r3080)~\cite{Bagnaschi:2011tu} & \PYTHIAV{8} & PDF4LHC15    & 48520             & N$^{3}$LO(QCD)+NLO(EW)           & Full       \\
                VBF                        & \POWHEGBOX (r3052)~\cite{Nason:2009ai}            & \PYTHIA     & PDF4LHC15    & 3780              & NNLO(QCD)+NLO(EW)                & Full       \\
                \WH                        & \POWHEGBOX (r3133)~\cite{Luisoni:2013kna}         & \PYTHIA     & PDF4LHC15    & 1370              & NNLO(QCD)+NLO(EW)                & Full       \\
                $q\bar{q} \rightarrow \ZH$ & \POWHEGBOX (r3133)~\cite{Luisoni:2013kna}         & \PYTHIAV{8} & PDF4LHC15    & 760               & NNLO(QCD)+NLO(EW                 & Full       \\
                \ttH                       & \MGMCatNLO                                        & \PYTHIAV{8} & NNPDF3.0     & 510               & NLO(QCD)+NLO(EW)                 & Full       \\
                $gg \rightarrow \ZH$       & \POWHEGBOX (r3133)                                & \PYTHIAV{8} & PDF4LHC15    & 120               & NLO+NLL(QCD)                     & Full       \\
                \bbH                       & \MGMCatNLO                                        & \PYTHIA     & CT10 NLO     & 490               & NNLO(5FS)+NLO(4FS)               & Full       \\
                t-channel \tH              & \MGMCatNLO                                        & \PYTHIAV{8} & CT10 NLO     & 70                & LO(4FS)                          & Full       \\
                $W$-associated \tH         & \MGMCatNLO                                        & \HERWIGpp   & CT10 NLO     & 20                & NLO(5FS)                         & Full       \\
                \bottomrule
            \end{tabular}
        }
    \end{center}
\end{table}


\section{Object and event selection}\label{sec:selection}
The photon selection and event selection for the present search follow those in another published ATLAS \hyy analysis~\cite{HIGG-2016-21}.
The subsections below detail the selection and identification of all detector-level objects used in the analysis, followed by the event selection criteria and the classification into signal and background control categories.

\subsection{Object selection}\label{sec:objectselection}
Photon candidates are reconstructed from energy clusters in the EM calorimeter~\cite{PERF-2013-05}.
The reconstruction algorithm searches for possible matches between energy clusters and tracks reconstructed in the inner detector and extrapolated to the calorimeter.
Well-reconstructed tracks matched to clusters are classified as electron candidates, while clusters without matching tracks are classified as unconverted photon candidates.
Clusters matched to a reconstructed conversion vertex or to pairs of tracks consistent with the hypothesis of a \yee conversion process are classified as converted photon candidates.
Photon energies are determined by summing the energies of all cells belonging to the associated cluster.
Simulation-based corrections are then applied to account for energy losses and leakage outside the cluster~\cite{PERF-2013-05}.
The absolute energy scale and response resolution is calibrated using $\Zboson \rightarrow e^+ e^-$ events from data.
For the photons considered in this analysis, the reconstruction efficiency for both the converted and unconverted photons is 97\%.
Photon identification is based on the lateral and longitudinal energy profiles of EM showers measured in the calorimeter~\cite{PERF-2013-04}.
The reconstructed photon candidates must satisfy tight photon identification criteria.
These exploit the fine granularity of the first layer of the EM calorimeter in order to reject background photons from hadron decays.
The photon identification efficiency varies as a function of \ET and $|\eta|$ and is typically 85--90\% (85--95\%) for unconverted (converted) photons in the range of $\SI{30}{\GeV} < \ET < \SI{100}{\GeV}$.

All photon candidates must satisfy a set of calorimeter- and track-based isolation criteria designed to reject the background from jets misidentified as photons and to maximise the signal significance of simulated \hyy events against the continuum background.
The calorimeter-based isolation variable \ETiso is defined as the sum of the energies of all topological clusters of calorimeter cells within $\Delta R = 0.2$ of the photon candidate, excluding clusters associated to the photon candidate.
The track-based isolation variable \pTiso is defined as the sum of the transverse momenta (\pT) of all tracks with $\pT > \SI{1}{\GeV}$ within $\Delta R = 0.2$ of the photon candidate, excluding tracks from photon conversions and tracks not associated with the interaction vertex.
Candidates with \ETiso larger than 6.5\% of their transverse energy or with \pTiso greater than 5\% of their transverse energy are rejected.
The efficiency of this isolation requirement is approximately 98\%.
Photons satisfying the isolation criteria are required to fall within the fiducial region of the EM calorimeter defined by \abseta $<$ 2.37, excluding a transition region between calorimeters ($1.37 < \abseta < 1.52$).
Among the photons satisfying the isolation and fiducial criteria, the two with the highest \pT are required to have $\ET/\myy > 0.35$ and $0.25$, where \myy is the invariant mass of the diphoton system.

A neural network, trained on a simulated gluon--gluon fusion single-Higgs-boson sample, is used to select the primary vertex most likely to have produced the diphoton pair.
The algorithm uses the directional information from the calorimeter and, in the case of converted photons, tracking information, to extrapolate the photon trajectories back to the beam axis.
Additionally, vertex properties such as the sum of the squared transverse momenta or the scalar sum of the transverse momenta of the tracks associated with the vertex, are used as inputs to this algorithm.
Due to the presence of two high-\pT jets in addition to the two photons, the efficiency for selecting the correct primary vertex is more than 85\%.
All relevant tracking and calorimetry variables are recalculated with respect to the chosen primary vertex~\cite{HIGG-2016-21}.

Jets are reconstructed via the FastJet package~\cite{Cacciari:2011ma} from topological clusters of energy deposits in calorimeter cells~\cite{PERF-2014-07}, using the \antikt algorithm~\cite{Cacciari:2008gp} with a radius parameter of $R = 0.4$.
Jets are corrected for contributions from pile-up by applying an event-by-event energy correction evaluated using calorimeter information~\cite{PERF-2014-03}.
They are then calibrated using a series of correction factors, derived from a mixture of simulated events and data, which correct for the different responses to EM and hadronic showers in each of the components of the calorimeters~\cite{PERF-2016-04}.
Jets that do not originate from the diphoton primary vertex, as detailed above, are rejected using the jet vertex tagger (JVT)~\cite{ATLAS-CONF-2014-018}, a multivariate likelihood constructed from two track-based variables.
A JVT requirement is applied to jets with $\SI{20}{\GeV} < \pT < \SI{60}{\GeV}$ and $\abseta < 2.4$. This requirement is 92\% efficient at selecting jets arising from the chosen primary vertex.
Jets are required to satisfy $\abseta < 2.5$ and $\pT > \SI{25}{\GeV}$; any jets among these that are within $\Delta R = 0.4$ of an isolated photon candidate or within $\Delta R = 0.2$ of an isolated electron candidate are discarded.

The selected jets are classified as \bjet{s} (those containing $b$-hadrons) or other jets using a multivariate classifier taking impact parameter information, reconstructed secondary vertex position and decay chain reconstruction as inputs~\cite{PERF-2012-04,ATL-PHYS-PUB-2015-022}.
Working points are defined by requiring the discriminant output to exceed a particular value that is chosen to provide a specific $b$-jet efficiency in an inclusive \ttbar sample.
Correction factors derived from \ttbar events with final states containing two leptons are applied to the simulated event samples to compensate for differences between data and simulation in the \btagging efficiency~\cite{PERF-2016-05}.
The analysis uses two working points which have a \btagging efficiency of 70\% (60\%), a $c$-jet rejection factor of 12 (35) and a light-jet rejection factor of 380 (1540) respectively.
Muons~\cite{PERF-2015-10} within $\Delta R = 0.4$ of a \btagged jet are used to correct for energy losses from semileptonic $b$-hadron decays.
This correction improves the energy measurement of \bjet{s} and improves the signal acceptance by 5--6\%.

\subsection{Event selection and categorisation}\label{sec:eventselection}
Events are selected for analysis if there are at least two photons and at least two jets, one or two of which are tagged as \bjet{s}, which satisfy the criteria outlined in \Sect{\ref{sec:objectselection}}.
The diphoton invariant mass is initially required to fall within a broad mass window of $\SI{105}{\GeV} < \myy < \SI{160}{\GeV}$.
In order to remain orthogonal to the ATLAS search for $\hh \rightarrow b\bar{b}b\bar{b}$~\cite{EXOT-2016-31}, any event with more than two \bjets using the 70\% efficient working point is rejected, before the remaining events are divided into three categories.
The 2-tag signal category consists of events with exactly two \bjets satisfying the requirement for the 70\% efficient working point.

Another signal category is defined using events failing this requirement but nevertheless containing exactly one \bjet identified using a more stringent (60\% efficient) working point.
Here the second jet, which is in this case not identified as a \bjet, is chosen using a boosted decision tree (BDT).
Different BDTs are used when applying the loose and tight kinematic selections.
These are optimised using simulated continuum background events as well as signal events from lower-mass or higher-mass resonances, respectively.
For each event in the simulated samples used for training the BDTs, every pair of jets in the event is considered.
A maximum of one of these jet pairs is correct; in the case of the background sample there are no correct pairs.
The training is then performed on these correct and incorrect jet pairs.
The BDTs use kinematic variables, namely jet \pT, dijet \pT, dijet mass, jet $\eta$, dijet $\eta$ and the $\Delta\eta$ between the selected jets, as well as information about whether each jet satisfied less stringent \btagging criteria.
The ranking of the jets from best to worst in terms of closest match between the dijet mass and \mH, highest jet \pT and highest dijet \pT are also used as inputs.
The jet with the highest BDT score is selected and the event is included in the 1-tag signal category.
The efficiency with which the correct jet is selected by this BDT is 60--80\% across the range of resonant and non-resonant signal hypotheses considered in this paper.
If the event contains no \bjet from either working point, the event is not directly used in the analysis, but is instead reserved for a 0-tag control category, which is used to provide data-driven estimates of the background shape in the signal categories.

Further requirements are then made on the \pT of the jets and on the mass of the dijet system, which differ for the loose and tight selections.
In the loose selection, the highest-\pT jet is required to have $\pT > \SI{40}{\GeV}$, and the next-highest-\pT jet must satisfy $\pT > \SI{25}{\GeV}$, with the invariant mass of the jet pair (\mjj) required to lie between \num{80} and \SI{140}{\GeV}.
For the tight selection, the highest-\pT and the next-highest-\pT jets are required to have $\pT > \SI{100}{\GeV}$ and $\pT > \SI{30}{\GeV}$, respectively, with $\SI{90}{\GeV} < \mjj < \SI{140}{\GeV}$.
Finally, in the resonant search, the diphoton invariant mass is required to be within \SI[parse-numbers=false]{4.7~(4.3)}{\GeV} of the Higgs boson mass for the loose (tight) selection.
This additional selection on \myy is optimised to contain at least 95\% of the simulated Higgs boson pair events for each mass hypothesis.

For non-resonant Higgs boson pair production, among events in the 2-tag category, the efficiency with which the kinematic requirements are satisfied is 10\% and 5.8\% for the loose and tight selections, respectively.
In the 1-tag category, the corresponding efficiencies are 7.2\% and 3.9\%, which are slightly lower than for the 2-tag category due to the lower probability of selecting the correct jet pair.
For the resonant analysis, efficiencies range from 6\% to 15.4\% in the 2-tag category and from 5.1\% to 12.3\% in the 1-tag category for $\SI{260}{\GeV} < \mX < \SI{1000}{\GeV}$.

Due to the differing jet kinematics, the signal acceptance is lower in all cases for the generated NLO signal than for a LO signal.
The acceptance of the LO prediction is approximately 15\% higher when using the tight selection and 10\% higher when using the loose selection.

\begin{figure}[!htb]
    \centering
    \subfloat[]{\includegraphics[width=0.48\textwidth]{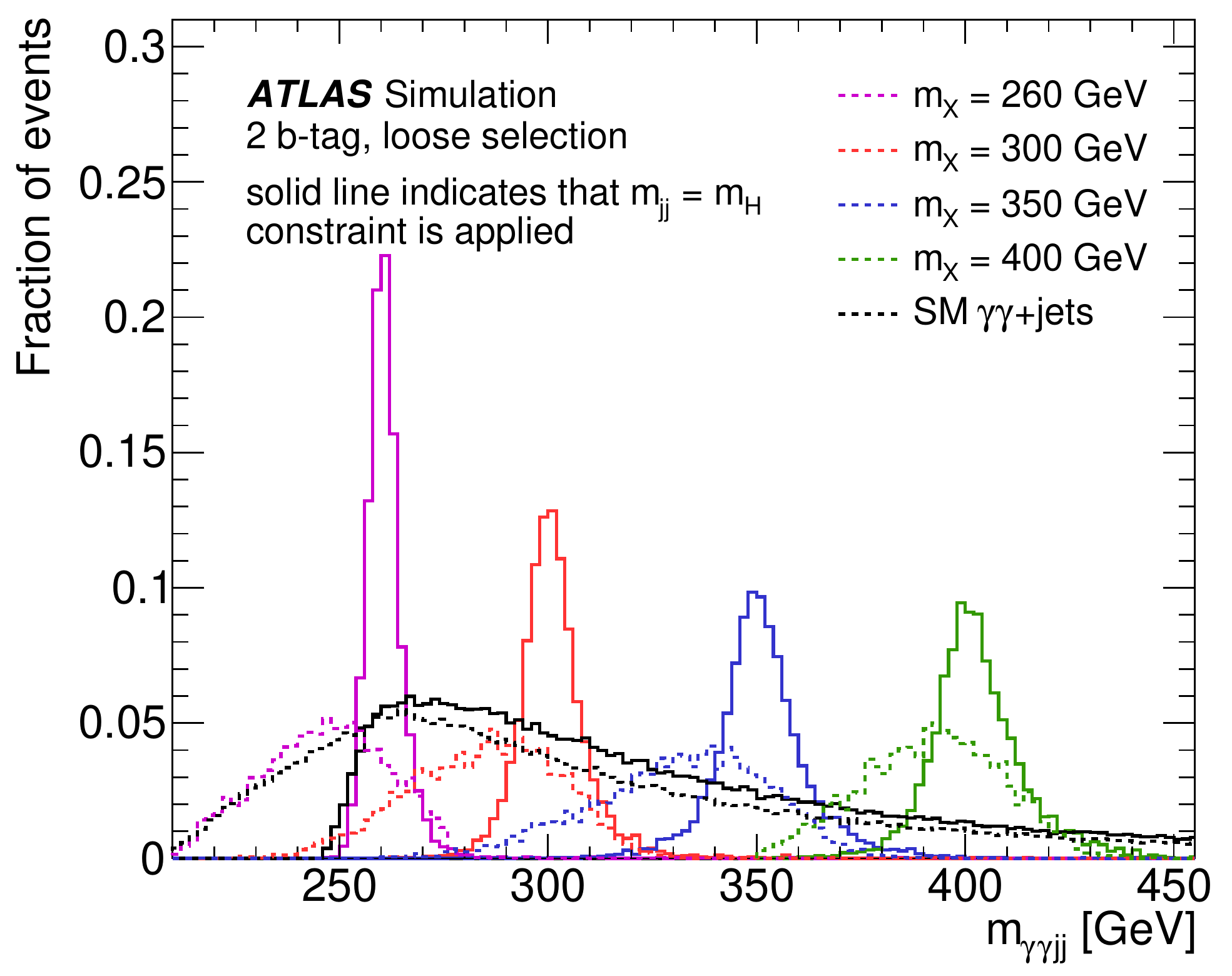}}
    \quad
    \subfloat[]{\includegraphics[width=0.48\textwidth]{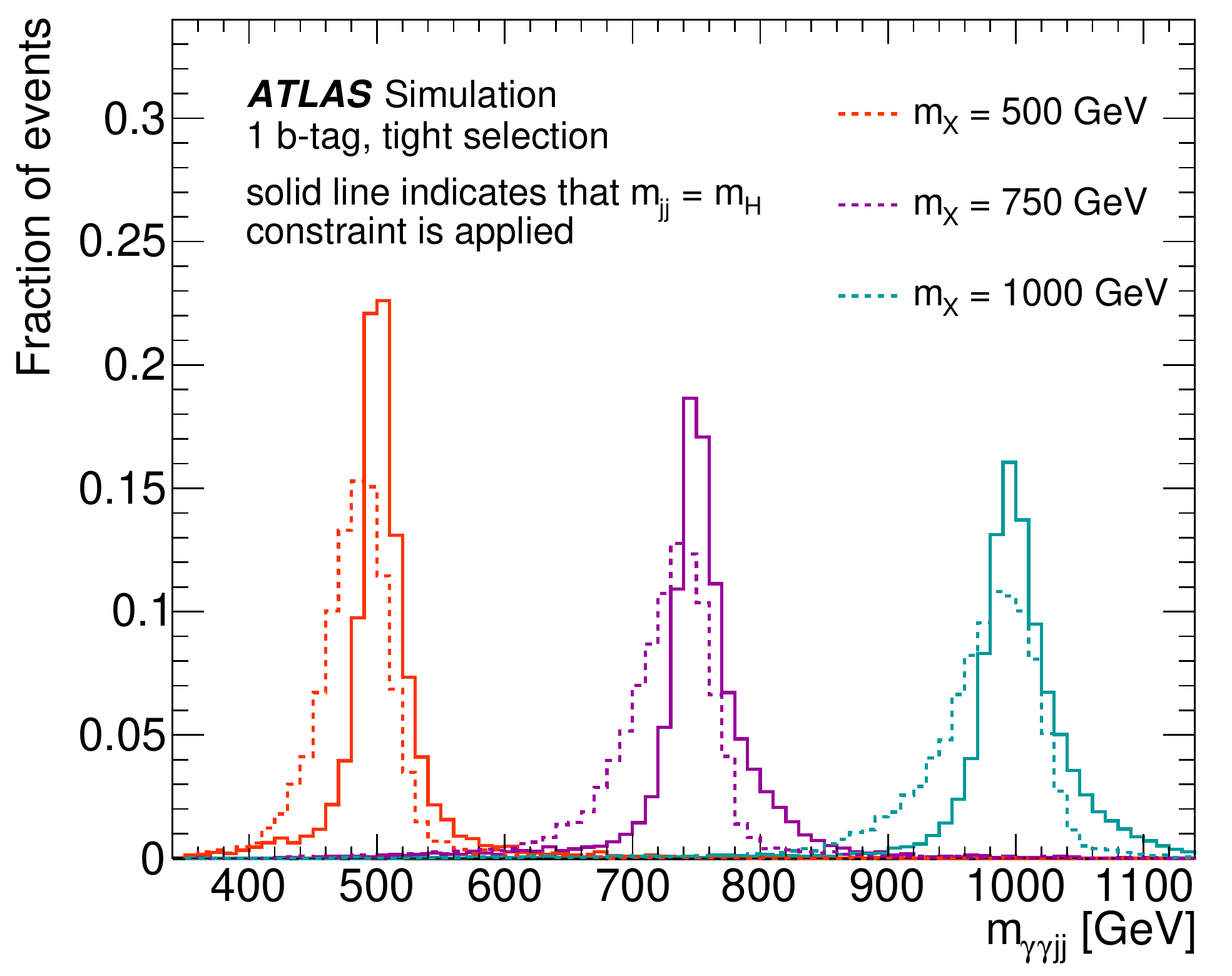}}
    \caption{Reconstructed \myyjj with (solid lines) and without (dashed lines) the dijet mass constraint, for a subset of the mass points used in the resonant analysis.
        The examples shown here are for (a) the 2-tag category with the loose selection and (b) the 1-tag category with the tight selection.
        The effect on the continuum background is also shown in (a).}
    \label{fig:mbb-scaling}
\end{figure}

In the resonant analysis, before reconstructing the four-object mass, \myyjj, the four-momentum of the dijet system is scaled by $\mH/\mjj$.
As shown in \Fig{\ref{fig:mbb-scaling}}, this improves the four-object mass resolution by $60$\% on average across the resonance mass range of interest.
It also modifies the shape of the non-resonant background in the region below \SI{270}{\GeV}.
After the correction, the \myyjj resolution is approximately 3\% for all signal hypotheses considered in this paper.
For the diphoton system, no such scaling is necessary due to the small (approximately $1.5$\%) diphoton mass resolution.

\section{Signal and background modelling}\label{sec:modelling}
Both the resonant and non-resonant searches for Higgs boson pairs proceed by performing unbinned maximum-likelihood fits to the data in the 1-tag and 2-tag signal categories simultaneously.
The non-resonant search involves a fit to the \myy distribution, while the search for resonant production uses the \myyjj distribution.
The signal-plus-background fit to the data uses parameterised forms for both the signal and background probability distributions.
These parameterised forms are determined through fits to simulated samples.

As the loose selection is used for resonances with $\mX \leq \SI{500}{\GeV}$ and the tight selection for resonances with $\mX \geq \SI{500}{\GeV}$, different ranges of \myyjj are used in each case.
For the loose (tight) selection, only events with \myyjj in the range $\SI{245}{\GeV} < \myyjj < \SI{610}{\GeV}$ ($\SI{335}{\GeV} < \myyjj < \SI{1140}{\GeV}$) are considered.
These ranges are the smallest that contain over 95\% of all of the simulated signal sample events with \mX below, or above, \SI{500}{\GeV} respectively.

\subsection{Background composition}
\label{sec:bkg_MC}
Contributions to the continuum diphoton background originate from \yy, \yj, \jy and \jj sources produced in association with jets, where $j$ denotes jets misidentified as photons and \yj and \jy differ by the jet faking the sub-leading or the leading photon candidate respectively.
These are determined from data using a double two-dimensional sideband method (2x2D) based on varying the photon identification and isolation criteria~\cite{STDM-2011-05,STDM-2010-08}.
The number and relative fraction of events from each of these sources is calculated separately for the 1- and 2-tag categories.
In each case the contribution from \yy events is in the range 80--90\%.

The choice of functional form used to fit the background in the final likelihood models is derived using simulated events.
Continuum \yy events were simulated using the \SHERPA event generator as described in \Sect{\ref{sec:data_MC}}.
As this prediction from \SHERPA does not provide a good description of the \myy spectrum in data, the mismodelling is corrected for using a data-driven reweighting function.

In the 0-tag control category, the number of events in data is high enough that the 2x2D method can be applied in bins of \myy.
The events generated by \SHERPA can also be divided into \yy, \yj, \jy and \jj sources based on the same photon identification and isolation criteria as used in data.
For each of these sources, the \myy distributions for both \SHERPA and the data are fit using an exponential function and the ratio of the two fit results is taken as an \myy-dependent correction function.
The size of the correction is less than 5\% for the majority of events.

These reweighting functions are then applied in the 1-tag and 2-tag signal categories to correct the shape of the \SHERPA prediction.
The fractional contribution from the different continuum background sources is fixed to the relative proportions derived in data with the 2x2D method.
Finally, the overall normalisation is chosen such that, in the disjoint sideband region $\SI{105}{\GeV} < \myy < \SI{120}{\GeV}$ and $\SI{130}{\GeV} < \myy < \SI{160}{\GeV}$, the total contribution from all backgrounds is equal to that from data.

\begin{figure}[!htb]
    \centering
    \subfloat[]{\includegraphics[width=0.48\textwidth]{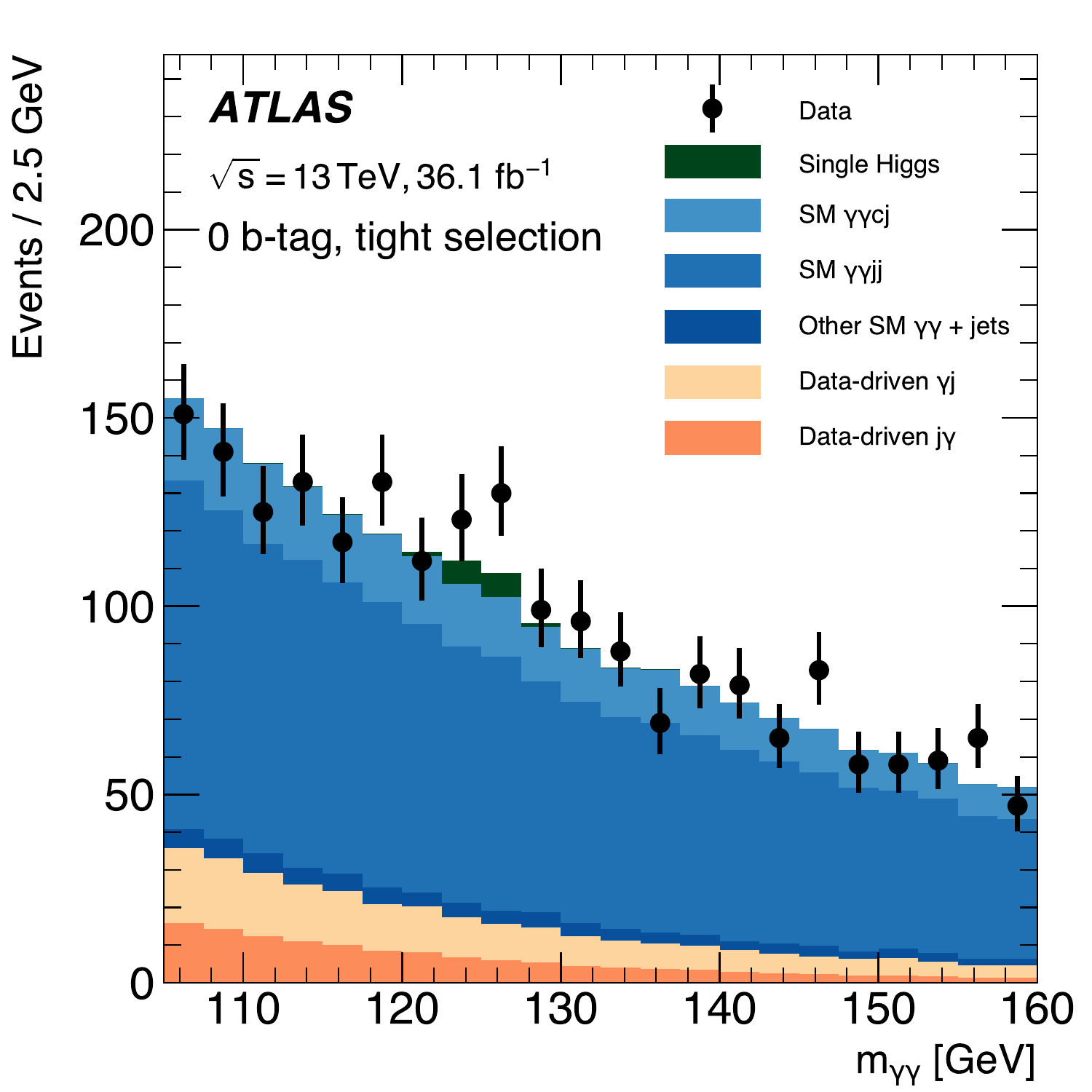}}
    \quad
    \subfloat[]{\includegraphics[width=0.48\textwidth]{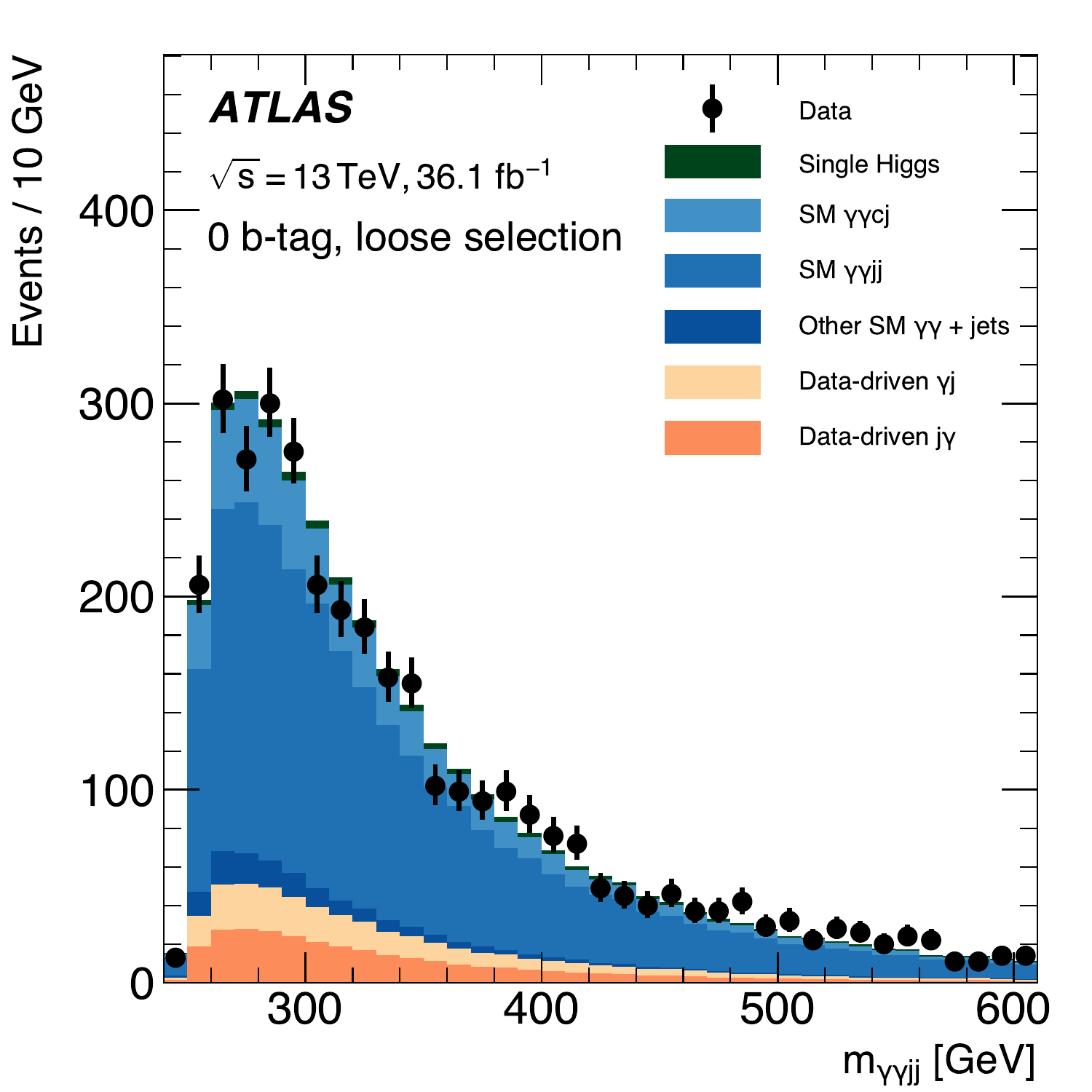}}
    \caption{The predicted number of background events from continuum diphoton plus jet production (blue), other continuum photon and jet production (orange) and single Higgs boson production (green) is compared with the observed data (black points) in the 0-tag control category for (a) the \myy distribution with the tight selection and (b) the \myyjj distribution with the loose selection.}
    \label{fig:modelling-background-decomposition}
\end{figure}

The contribution from \yy produced in association with jets is further divided in accord with the flavours of the two jets (for example $bb$, $bc$, $c+\text{light jet}$).
This decomposition is taken directly from the proportions predicted by the \SHERPA event generator and no attempt is made to classify the data according to jet flavour.
The continuum background in the 1-tag category comes primarily from $\yy bj$ events (${\sim}60\%$) and in the 2-tag category from $\yy bb$ events (${\sim}80\%$).
A comparison between data in the 0-tag control category and this data-driven prediction of the total background can be seen in \Fig{\ref{fig:modelling-background-decomposition}(a)} for the \myy distribution from the tight selection and in \Fig{\ref{fig:modelling-background-decomposition}(b)} for the \myyjj distribution from the loose selection.

\subsection{Signal modelling for the non-resonant analysis}
\label{subsec:signal_modelling_nonres}
The shape of the diphoton mass distribution in \hhyyjj events is described by the double-sided Crystal Ball function~\cite{HIGG-2016-21}, consisting of a Gaussian core with power-law tails on either side.
The parameters of this model are determined through fits to the simulated non-resonant SM \hh sample described in \Sect{\ref{subsec:data_MC_samples}}.

\subsection{Background modelling for the non-resonant analysis}
\label{subsec:bg_modelling_nonres}
For the non-resonant analysis, the continuum \myy background is modelled using a functional form obtained from a fit to the data.
The potential bias arising from this procedure, termed `spurious signal', is estimated by performing signal-plus-background fits to the combined continuum background from simulation, including the \yy, \yj, \jy and \jj components~\cite{HIGG-2016-21}.
The maximum absolute value of the extracted signal, for a signal in the range $\SI{121}{\GeV} < \myy < \SI{129}{\GeV}$, is taken as the bias.
This method is used to discriminate between different potential fit functions -- the function chosen is the one with the smallest spurious signal bias.
If multiple functions have the same bias, the one with the smallest number of parameters is chosen.
The first-order exponential function has the smallest bias among the seven functions considered and is therefore chosen.
The background from single Higgs boson production is described using a double-sided Crystal Ball function, with its parameters determined through fits to the appropriate simulated samples.

\subsection{Signal modelling for the resonant analysis}
For each resonant hypothesis, a fit is performed to the \myyjj distribution of the simulated events in a window around the nominal \mX.
The shape of this distribution is described using a function consisting of a Gaussian core with exponential tails on either side.
A simultaneous fit to all signal samples is carried out in which each of the model parameters is further parameterised in terms of \mX.
This allows the model to provide a prediction for any mass satisfying $\SI{260}{\GeV} < \mX < \SI{1000}{\GeV}$, where these boundaries reflect the smallest and largest \mX values among the generated samples described in \Sect{\ref{subsec:data_MC_samples}}.

\subsection{Background modelling for the resonant analysis}
\label{subsec:bg_modelling_res}
For the resonant analysis, a spurious-signal study is also carried out, using the \myyjj distribution for events within the \myy window described in \Sect{\ref{sec:eventselection}}.
The background used to evaluate the spurious-signal contribution is a combination of the continuum \myy backgrounds together with the single Higgs boson backgrounds.

Due to the different \myyjj ranges used with the loose and tight selections, the shape of the \myyjj distribution differs between these two cases and hence different background functions are considered.
For the loose (tight) mass selection, the Novosibirsk function\footnote{$P(x) = \mathrm{e}^{-0.5^{(\ln q_{y})^{2}/\Lambda^{2} + \Lambda^{2}}}$ where $q_{y} = 1 + \Lambda(x - x_{0})/\sigma \times \frac{\sinh(\Lambda\sqrt{\ln{4}})}{\Lambda\sqrt{\ln{4}}}$~\cite{Ikeda:1999aq}.} (exponential function) has the smallest bias among the three (four) functions considered and is therefore chosen.
As a result, for low-mass resonances both the signal and background fit functions have a characteristic peaked shape.
This degeneracy could potentially introduce a bias in the extracted signal \xs.
In order to stabilise the background fit, nominal values of the shape parameters are estimated by fitting to the simulated events described in \Sect{\ref{sec:bkg_MC}}.
The shape is then allowed to vary in the likelihood to within the statistical covariance of this template fit.
Experimental systematics on the background shape have a small effect and are neglected.
The normalisation of the background is estimated by interpolating the \myy sideband data.
Additionally, a simple bias test is performed by drawing pseudo-data sets from the overall probability distribution created by combining the Novosibirsk background function with the signal function.
For each mass point and each value of the injected signal \xs, fits are performed on the ensemble of pseudo-data sets and the median extracted signal \xs is recorded.
For resonances with masses below \SI{400}{\GeV}, a small correction is applied to remove the observed bias.
The correction is less than \SI[parse-numbers=false]{\pm0.05}{\pico\barn} everywhere and a corresponding uncertainty of \SI[parse-numbers=false]{\pm0.02}{\pico\barn} in this correction is applied to the extracted signal \xs.
The corresponding uncertainty in the number of events in each category is roughly half that of the spurious signal.


\section{Systematic uncertainties}\label{sec:systematics}
Although statistical uncertainties dominate the sensitivity of this analysis given the small number of events, care is taken to make the best possible estimates of all systematic uncertainties, as described in more detail below.

\subsection{Theoretical uncertainties}
Theoretical uncertainties in the production \xs of single Higgs bosons are estimated by varying the renormalisation and factorisation scales.
In addition, uncertainties due to the PDF and the running of the QCD coupling constant ($\alphas$) are considered.
The scale uncertainties reach a maximum of $^{+20\%}_{-24\%}$ and the PDF+$\alphas$ uncertainty is not more than $\pm$3.6\%~\cite{deFlorian:2016spz}.
An uncertainty in the rate of Higgs boson production with associated heavy-flavour jets is also considered.
A 100\% uncertainty is assigned to the \ggH and \WH production modes, motivated by studies of heavy-flavour production in association with top-quark pairs~\cite{TOPQ-2012-16} and $W$ boson production in association with $b$-jets~\cite{STDM-2012-11}.
No heavy-flavour uncertainty is assigned to the \ZH and \ttH production modes, where the dominant heavy-flavour contribution is already accounted for in the LO process.
Finally, additional theoretical uncertainties in single Higgs boson production from uncertainties in the \hyy and \hbb branching fractions are $^{+2.9\%}_{-2.8\%}$ and $\pm$1.7\%, respectively~\cite{deFlorian:2016spz}.

The same sources of uncertainty are considered on the SM \hh signal samples.
The effect of scale and PDF+$\alphas$ uncertainties on the NNLO \xs for SM Higgs boson pair production are 4--8\% and 2--3\% respectively.
In addition, an uncertainty of 5\% arising from the use of infinite top-quark mass in the EFT approximation is taken into account~\cite{Borowka:2016ehy}.
The \xs, scale and PDF uncertainties are decorrelated between the single Higgs boson background and SM \hh signal.

In the search for resonant Higgs boson pair production, uncertainties arising from scale and PDF uncertainties, which primarily affect the signal yield, are neglected.
For this search, the SM non-resonant \hh production is considered as a background, with an overall uncertainty on the \xs of $^{+7\%}_{-8\%}$.
Interference between SM \hh and the BSM signal is neglected.

For all samples, systematic differences between alternative models of parton showering and hadronisation were considered and found to have a negligible impact.

\subsection{Experimental uncertainties}
The systematic uncertainty in the integrated luminosity for the data in this analysis is 2.1\%.
It is derived following a methodology similar to that detailed in Ref.~\cite{DAPR-2013-01}, using beam-separation scans performed in 2015 and 2016.

The efficiency of the diphoton trigger is measured using bootstrap methods \cite{TRIG-2016-01}, and is found to be 99.4\% with a systematic uncertainty of 0.4\%.
Uncertainties associated with the vertex selection algorithm have a negligible impact on the signal selection efficiency.

Differences between data and simulation give rise to uncertainties in the calibration of the photons and jets used in this analysis.
As the continuum backgrounds are estimated from data, these uncertainties are applied only to the signal processes and to the single-Higgs-boson background process.
Experimental uncertainties are propagated through the full analysis procedure, including the kinematic and BDT selections.
The relevant observables are then constructed, before signal and background fits are performed as described in \Sect{\ref{sec:modelling}}.
Changes in the peak location ($m_\mathrm{peak}$), width ($\sigma_\mathrm{peak}$) and expected yield in \myy (\myyjj) for the non-resonant (resonant) model, relative to the nominal fits, are extracted.
The tail parameters are kept at their nominal values in these modified fits.
For the resonant analysis, systematic uncertainties are evaluated for each \mX and the maximum across the range is taken as a conservative uncertainty.

The dominant yield uncertainties are listed in \Tab{\ref{tab:systematic_uncertainties}}.
Uncertainties in the photon identification and isolation directly affect the diphoton selection efficiency; jet energy scale and resolution uncertainties affect the $m_{bb}$ window acceptance \cite{ATL-PHYS-PUB-2015-015,PERF-2012-01,PERF-2016-04}, while flavour-tagging uncertainties lead to migration of events between categories.
Uncertainties in the peak location (width), which are mainly due to uncertainties in the photon energy scale (energy resolution), are about 0.2--0.6\% (5--14\%) for both the single-Higgs-boson and Higgs boson pair samples in the resonant and non-resonant analyses.

The spurious signal for the chosen background model, as defined in \Sects{\ref{subsec:bg_modelling_nonres}}{\ref{subsec:bg_modelling_res}}, is assessed as an additional uncertainty in the total number of signal events in each category.
In the 2-tag (1-tag) category, the uncertainty corresponds to 0.63 (0.25) events for the non-resonant analysis, 0.58 (2.06) events for the resonant analysis with the loose selection, and 0.21 (0.89) events for the resonant analysis with the tight selection.

Finally, as described in \Sect{\ref{subsec:bg_modelling_res}}, an \mX-dependent correction to the signal \xs, together with its associated uncertainty, is applied in the case of the resonant analysis at low masses to adjust for a small degeneracy bias.

\begin{table}[!htb]
    \begin{center}
    \caption{Summary of dominant systematic uncertainties affecting expected yields in the resonant and non-resonant analyses.
        For the non-resonant analysis, uncertainties in the Higgs boson pair signal and SM single-Higgs-boson backgrounds are presented.
        For the resonant analysis, uncertainties on the Higgs boson pair signal for the loose and tight selections are presented.
        Sources marked `-' and other sources not listed in the table are negligible by comparison.
        No systematic uncertainties related to the continuum background are considered, since this is derived through a fit to the observed data.}
    \label{tab:systematic_uncertainties}

        \resizebox{\textwidth}{!}
        {\normalsize
            \begin{tabular}{l l r@{} @{}D{.}{.}{1.1} r@{} @{}r@{} @{}D{.}{.}{1.1}@{} @{}l r@{} @{}D{.}{.}{1.1} r@{} @{}r@{} @{}D{.}{.}{2.1}@{} @{}l r@{} @{}D{.}{.}{1.1} r@{} @{}r@{} @{}D{.}{.}{1.1}@{} @{}l r@{} D{.}{.}{1.1} r@{} @{}r@{} @{}D{.}{.}{1.1}@{} @{}l}
                    \toprule
                \multicolumn{2}{c}{\multirow{2}{*}{Source of systematic uncertainty}}                                 & \multicolumn{24}{c}{\% effect relative to nominal in the 2-tag (1-tag) category}                                                                   \\
                                                         &                                                            & \multicolumn{12}{c}{Non-resonant analysis}                             & \multicolumn{12}{c}{Resonant analysis: BSM \hh}                           \\
                \midrule
                                                         &                                                            & \multicolumn{6}{c}{SM \hh signal} & \multicolumn{6}{c}{Single-$H$ bkg} & \multicolumn{6}{c}{Loose selection} & \multicolumn{6}{c}{Tight selection} \\
                \midrule
                \multicolumn{2}{l}{Luminosity}                                                                        & $\pm$ & 2.1 & ( & $\pm$ & 2.1 & ) & $\pm$ & 2.1 & ( & $\pm$ & 2.1  & ) & $\pm$ & 2.1 & ( & $\pm$ & 2.1 & )   & $\pm$ & 2.1 & ( & $\pm$ & 2.1 & )   \\
                \multicolumn{2}{l}{Trigger}                                                                           & $\pm$ & 0.4 & ( & $\pm$ & 0.4 & ) & $\pm$ & 0.4 & ( & $\pm$ & 0.4  & ) & $\pm$ & 0.4 & ( & $\pm$ & 0.4 & )   & $\pm$ & 0.4 & ( & $\pm$ & 0.4 & )   \\
                \multicolumn{2}{l}{Pile-up modelling}                                                                 & $\pm$ & 3.2 & ( & $\pm$ & 1.3 & ) & $\pm$ & 2.0 & ( & $\pm$ & 0.8  & ) & $\pm$ & 4.0 & ( & $\pm$ & 4.2 & )   & $\pm$ & 4.0 & ( & $\pm$ & 3.8 & )   \\
                \midrule
                \multirow{4}{*}{Photon}                  & identification                                             & $\pm$ & 2.5 & ( & $\pm$ & 2.4 & ) & $\pm$ & 1.7 & ( & $\pm$ & 1.8  & ) & $\pm$ & 2.6 & ( & $\pm$ & 2.6 & )   & $\pm$ & 2.5 & ( & $\pm$ & 2.5 & )   \\
                                                         & isolation                                                  & $\pm$ & 0.8 & ( & $\pm$ & 0.8 & ) & $\pm$ & 0.8 & ( & $\pm$ & 0.8  & ) & $\pm$ & 0.8 & ( & $\pm$ & 0.8 & )   & $\pm$ & 0.9 & ( & $\pm$ & 0.9 & )   \\
                                                         & energy resolution                                          & \multicolumn{6}{c}{-}             & \multicolumn{6}{c}{-}              & $\pm$ & 1.0 & ( & $\pm$ & 1.3 & )   & $\pm$ & 1.8 & ( & $\pm$ & 1.2 & )   \\
                                                         & energy scale                                               & \multicolumn{6}{c}{-}             & \multicolumn{6}{c}{-}              & $\pm$ & 0.9 & ( & $\pm$ & 3.0 & )   & $\pm$ & 0.9 & ( & $\pm$ & 2.4 & )   \\
                \midrule
                \multirow{2}{*}{Jet}                     & energy resolution                                          & $\pm$ & 1.5 & ( & $\pm$ & 2.2 & ) & $\pm$ & 2.9 & ( & $\pm$ & 6.4  & ) & $\pm$ & 7.5 & ( & $\pm$ & 8.5 & )   & $\pm$ & 6.4 & ( & $\pm$ & 6.4 & )   \\
                                                         & energy scale                                               & $\pm$ & 2.9 & ( & $\pm$ & 2.7 & ) & $\pm$ & 7.8 & ( & $\pm$ & 5.6  & ) & $\pm$ & 3.0 & ( & $\pm$ & 3.3 & )   & $\pm$ & 2.3 & ( & $\pm$ & 3.4 & )   \\
                \midrule
                \multirow{3}{*}{Flavour tagging}         & $b$-jets                                                   & $\pm$ & 2.4 & ( & $\pm$ & 2.5 & ) & $\pm$ & 2.3 & ( & $\pm$ & 1.4  & ) & $\pm$ & 3.4 & ( & $\pm$ & 2.6 & )   & $\pm$ & 2.5 & ( & $\pm$ & 2.6 & )   \\
                                                         & $c$-jets                                                   & $\pm$ & 0.1 & ( & $\pm$ & 1.0 & ) & $\pm$ & 1.8 & ( & $\pm$ & 11.6 & ) & \multicolumn{6}{c}{-}               & \multicolumn{6}{c}{-}               \\
                                                         & light-jets                                                 & $<$   & 0.1 & ( & $\pm$ & 5.0 & ) & $\pm$ & 1.6 & ( & $\pm$ & 2.2  & ) & \multicolumn{6}{c}{-}               & \multicolumn{6}{c}{-}               \\
                \midrule
                \multirow{4}{*}{Theory}                  & PDF{+}$\alphas$                                            & $\pm$ & 2.3 & ( & $\pm$ & 2.3 & ) & $\pm$ & 3.1 & ( & $\pm$ & 3.3  & ) & \multicolumn{6}{c}{n/a}             & \multicolumn{6}{c}{n/a}             \\
                                                         & \multirow{2}{*}{Scale}                                     & $+$   & 4.3 & ( & $+$   & 4.3 & ) & $+$   & 4.9 & ( & $+$   & 5.3  & ) & \multicolumn{6}{c}{n/a}             & \multicolumn{6}{c}{n/a}             \\
                                                         &                                                            & $-$   & 6.0 & ( & $-$   & 6.0 & ) & $+$   & 7.0 & ( & $+$   & 8.0  & ) & \multicolumn{6}{c}{n/a}             & \multicolumn{6}{c}{n/a}             \\
                                                         & EFT                                                        & $\pm$ & 5.0 & ( & $\pm$ & 5.0 & ) & \multicolumn{6}{c}{n/a}            & \multicolumn{6}{c}{n/a}             & \multicolumn{6}{c}{n/a}             \\
                \bottomrule
            \end{tabular}
        }
    \end{center}
\end{table}


\section{Results}
\label{sec:results}
The observed data are in good agreement with the data-driven background expectation, as summarised in \Tab{\ref{tab:results-nEvents-observed-expected}}.
Across all categories, the number of observed events in data is compatible with the number of expected background events within the calculated uncertainties.

\begin{table}[!htb]
    \begin{center}
        \caption{Expected and observed numbers of events in the 1-tag and 2-tag categories for events passing the selection for the resonant analysis, including the \myy requirement.
        The event numbers quoted for the SM Higgs boson pair signal assume that the total production \xs is \diHiggsXS.
        The uncertainties on the continuum background are those arising from the fitting procedure.
        The uncertainties on the single-Higgs-boson and Higgs boson pair backgrounds are from Monte Carlo statistical error.
        The loose and tight selections are not orthogonal.}
        \label{tab:results-nEvents-observed-expected}
        \sisetup{separate-uncertainty, table-format = 1.3(4), table-align-exponent = false, table-align-uncertainty = true}
        \resizebox{\textwidth}{!}
        {\normalsize
        \begin{tabular}{l D{.}{.}{3.3} @{}c@{} D{.}{.}{1.3} D{.}{.}{2.3} @{}c@{} D{.}{.}{1.3} D{.}{.}{2.3} @{}c@{} D{.}{.}{1.3} D{.}{.}{1.3} @{}c@{} D{.}{.}{1.3}}
            \toprule
                                             & \multicolumn{6}{c}{1-tag}                                                  & \multicolumn{6}{c}{2-tag}                                                 \\
                                             & \multicolumn{3}{c}{Loose selection}  & \multicolumn{3}{c}{Tight selection} & \multicolumn{3}{c}{Loose selection} & \multicolumn{3}{c}{Tight selection} \\
            \midrule
            Continuum background             & 117.5 & $\pm$ & 4.7                  & 15.7  & $\pm$ & 1.6                 & 21.0  & $\pm$ & 2.0                 & 3.74  & $\pm$ & 0.78                \\
            SM single-Higgs-boson background & 5.51  & $\pm$ & 0.10                 & 2.20  & $\pm$ & 0.05                & 1.63  & $\pm$ & 0.04                & 0.56  & $\pm$ & 0.02                \\
            \midrule
            Total background                 & 123.0 & $\pm$ & 4.7                  & 17.9  & $\pm$ & 1.6                 & 22.6  & $\pm$ & 2.0                 & 4.30  & $\pm$ & 0.79                \\
            \midrule
            SM Higgs boson pair signal       & 0.219 & $\pm$ & 0.006                & 0.120 & $\pm$ & 0.004               & 0.305 & $\pm$ & 0.007               & 0.175 & $\pm$ & 0.005               \\
            \midrule
            Data                             & 125   &       &                      & 19    &       &                     & 21    &       &                     & 3     &       &                     \\
            \bottomrule
        \end{tabular}
        }
    \end{center}
\end{table}

The signal and background models described in \Sect{\ref{sec:modelling}} are used to construct an unbinned likelihood function which is maximised with respect to the observed data.
In each case the parameter of interest is the signal \xs, which is related in the likelihood model to the number of signal events after considering the integrated luminosity, branching ratio, phase-space acceptance and detection efficiency of the respective categories.
The models for the statistically independent 1-tag and 2-tag categories are simultaneously fit to the data by maximizing the product of their likelihoods.
The likelihood model also includes a number of nuisance parameters associated with the background shape and normalisation, as well as the theoretical and experimental systematic uncertainties described in \Sect{\ref{sec:systematics}}.
These nuisance parameters are included in the likelihood as terms which modulate their respective parameters, such as signal yield, along with a constraint term which encodes the scale of the uncertainty by reducing the likelihood when the parameter is pulled from its nominal value.
In general the nuisance parameter for each systematic uncertainty has a correlated effect between 1-tag and 2-tag categories, with the exception of the spurious signal and background shape parameters, which are considered as individual degrees of freedom in each category.

\begin{figure}[!htb]
    \subfloat[]{\includegraphics[width=0.48\textwidth]{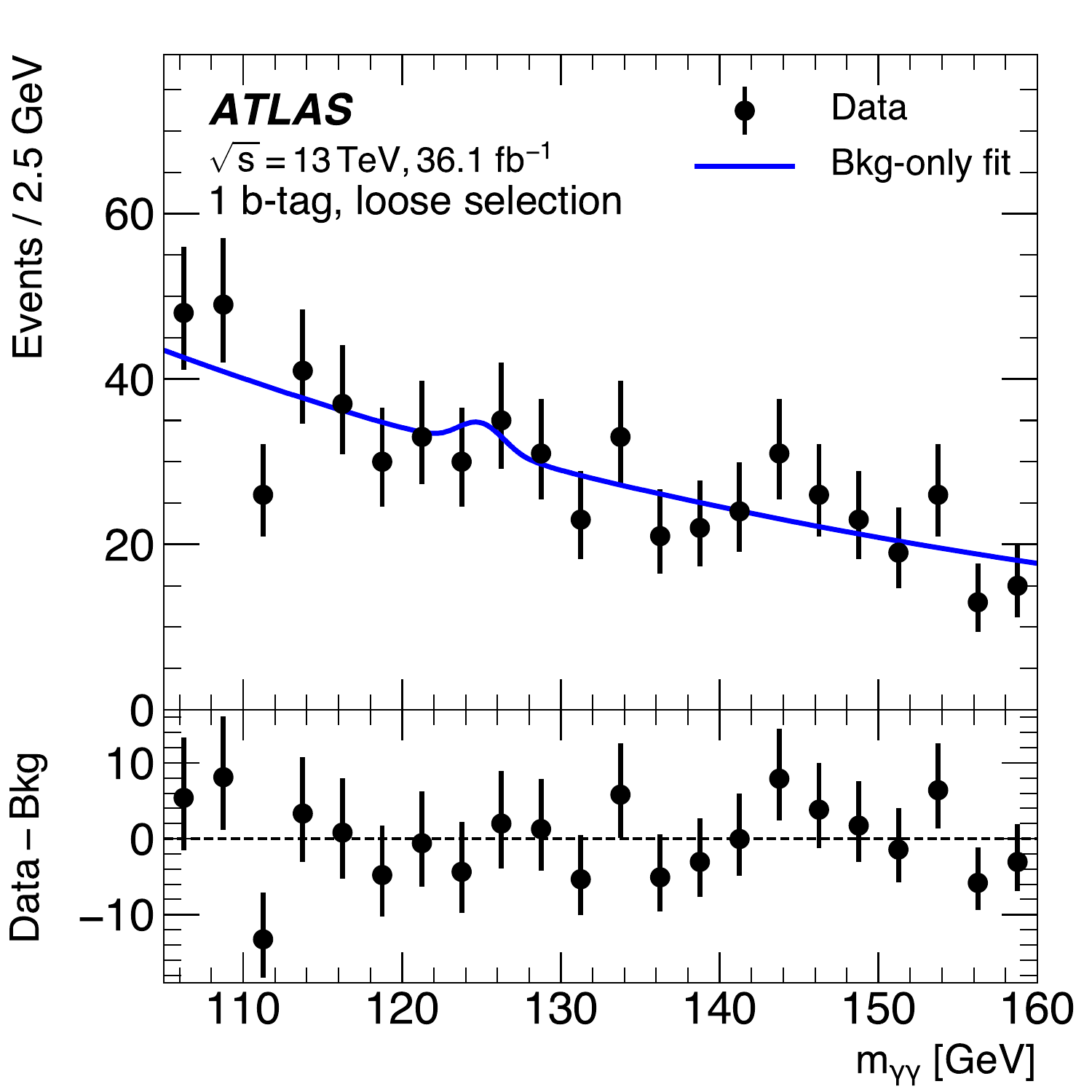}}
    \quad
    \subfloat[]{\includegraphics[width=0.48\textwidth]{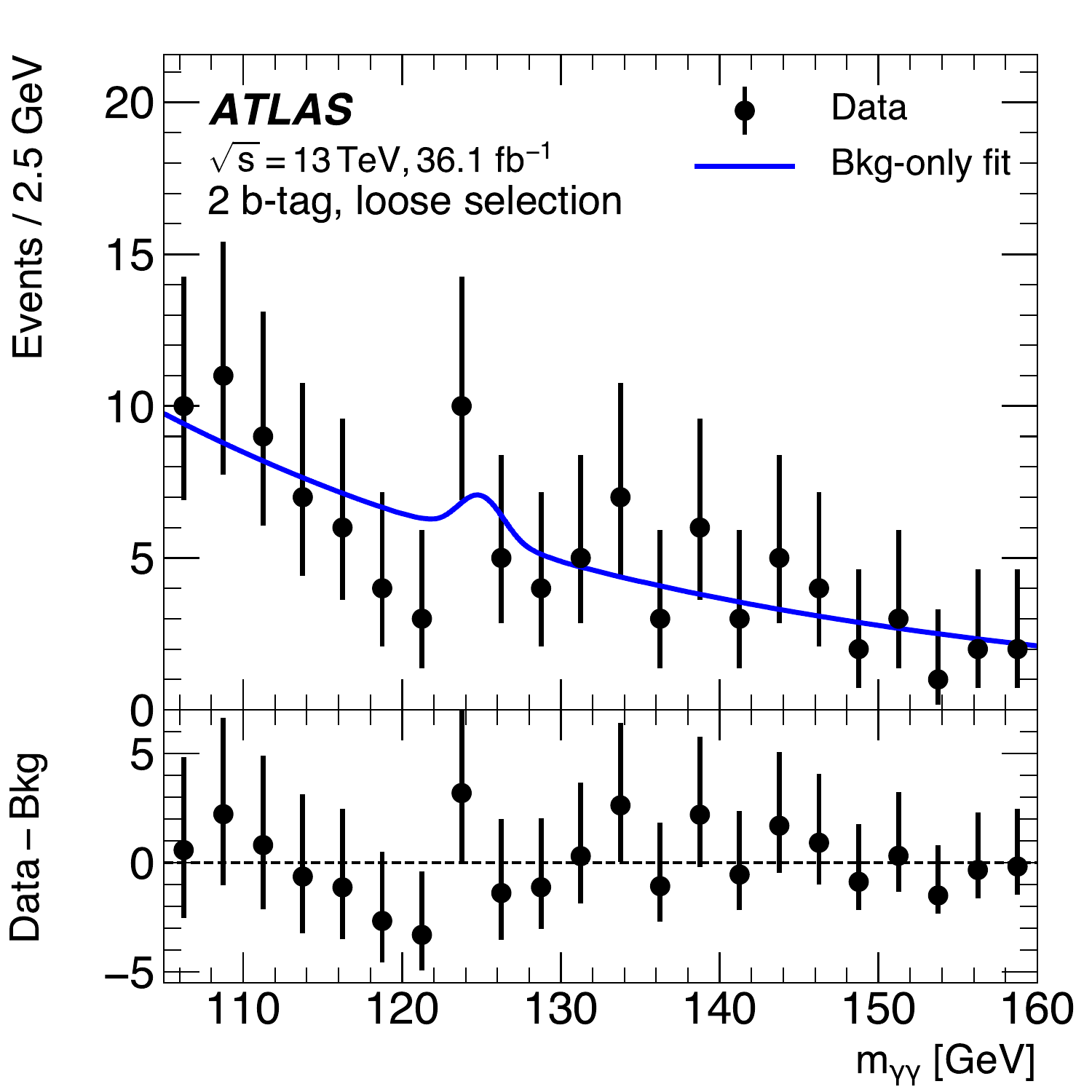}}
    \\
    \subfloat[]{\includegraphics[width=0.48\textwidth]{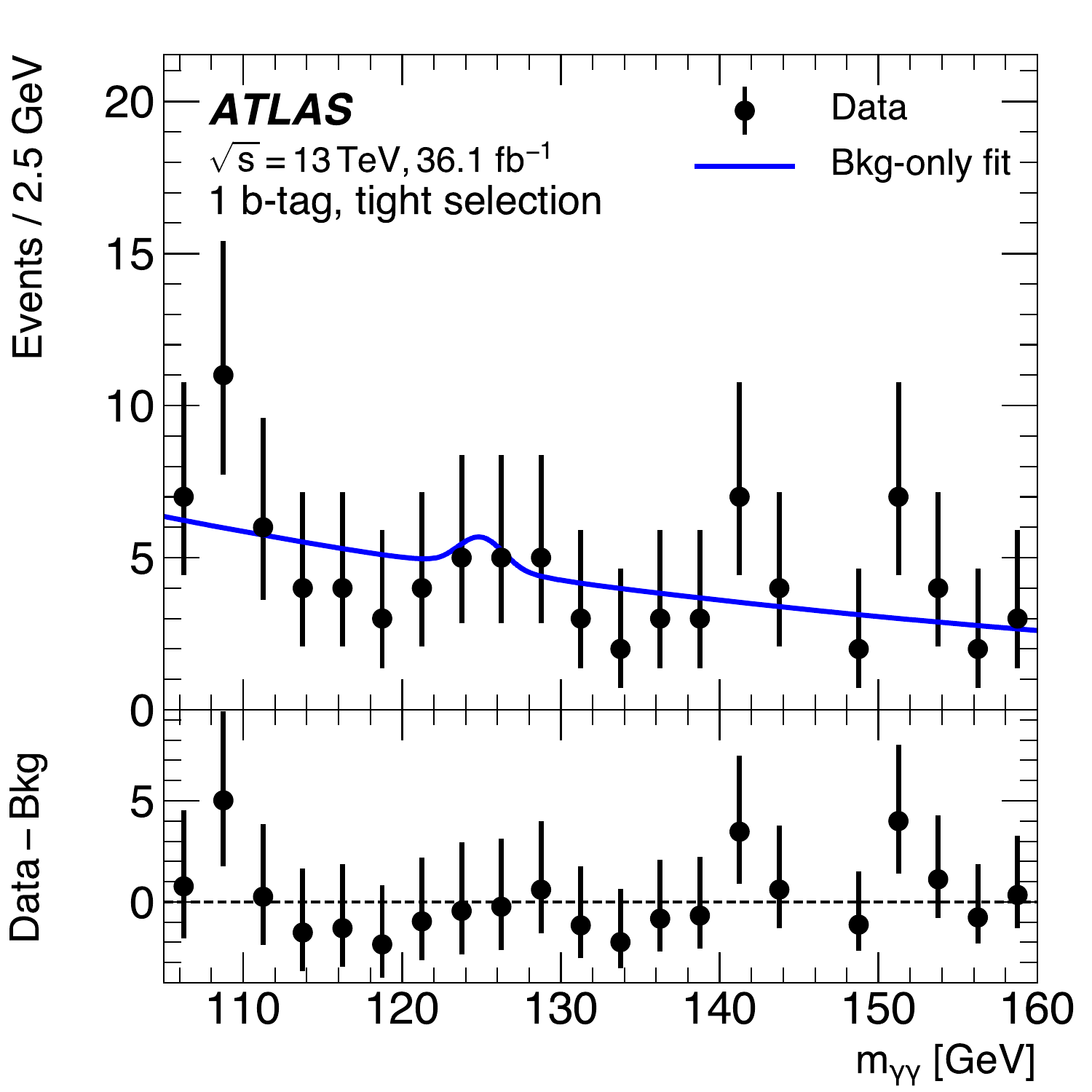}}
    \quad
    \subfloat[]{\includegraphics[width=0.48\textwidth]{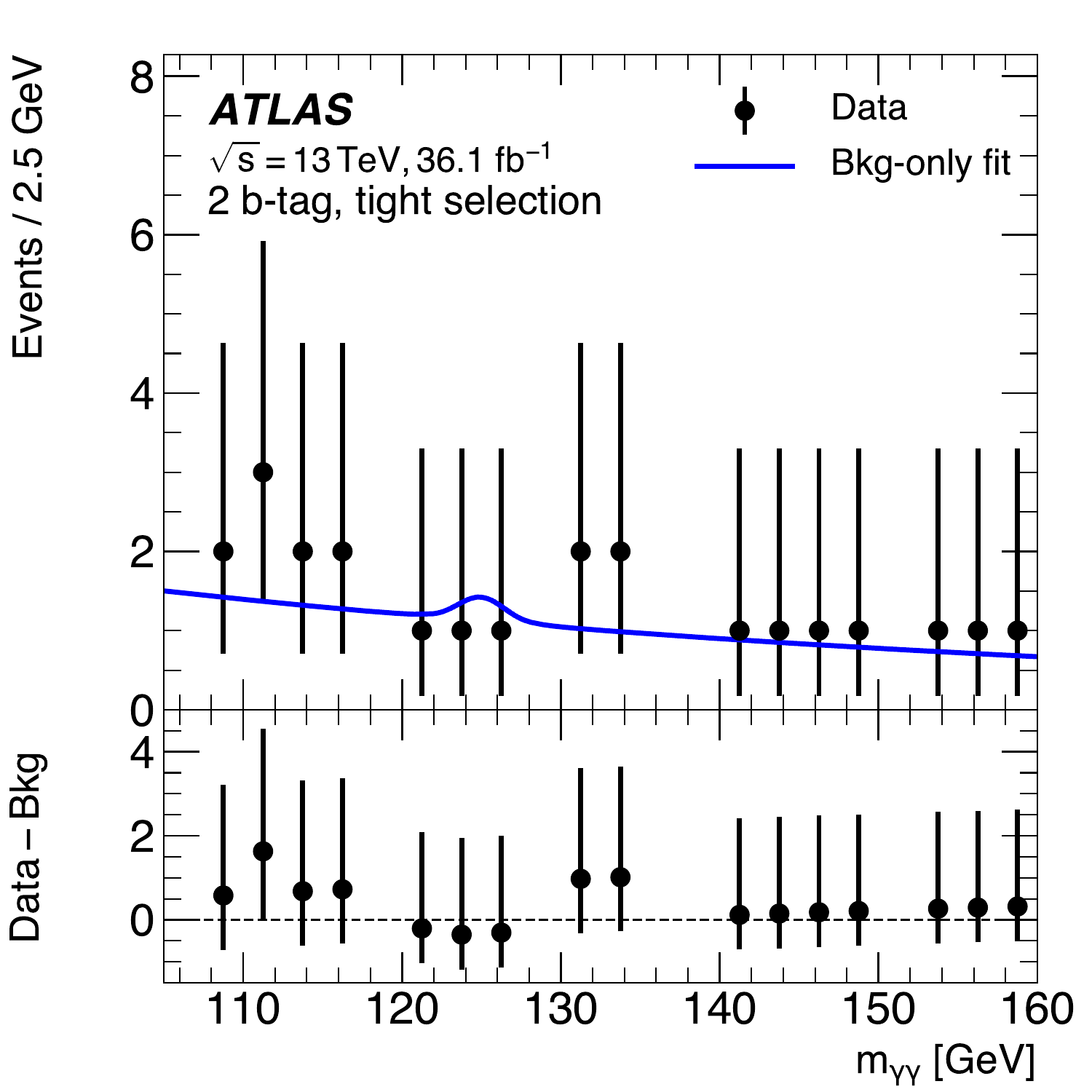}}
    \caption{For the non-resonant analysis, data (black points) are compared with the background-only fit (blue solid line) for \myy in the 1-tag (left) and 2-tag (right) categories with the loose (top) and tight (bottom) selections.
        Both the continuum \yy background and the background from single Higgs boson production are considered.
        The lower panel shows the residuals between the data and the best-fit background.
        }
    \label{results-myy}
\end{figure}

\begin{figure}[!htb]
    \subfloat[]{\includegraphics[width=0.48\textwidth]{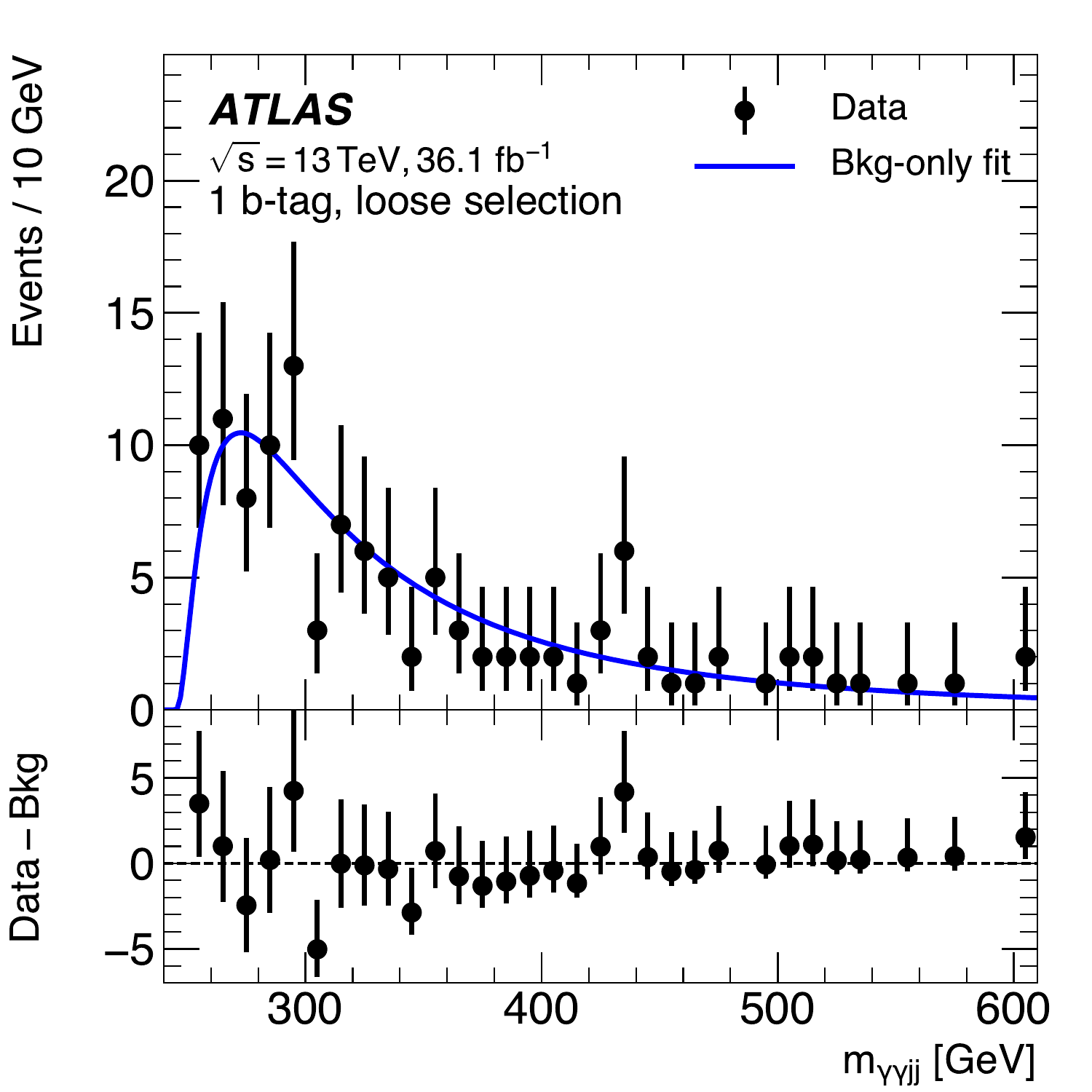}}
    \quad
    \subfloat[]{\includegraphics[width=0.48\textwidth]{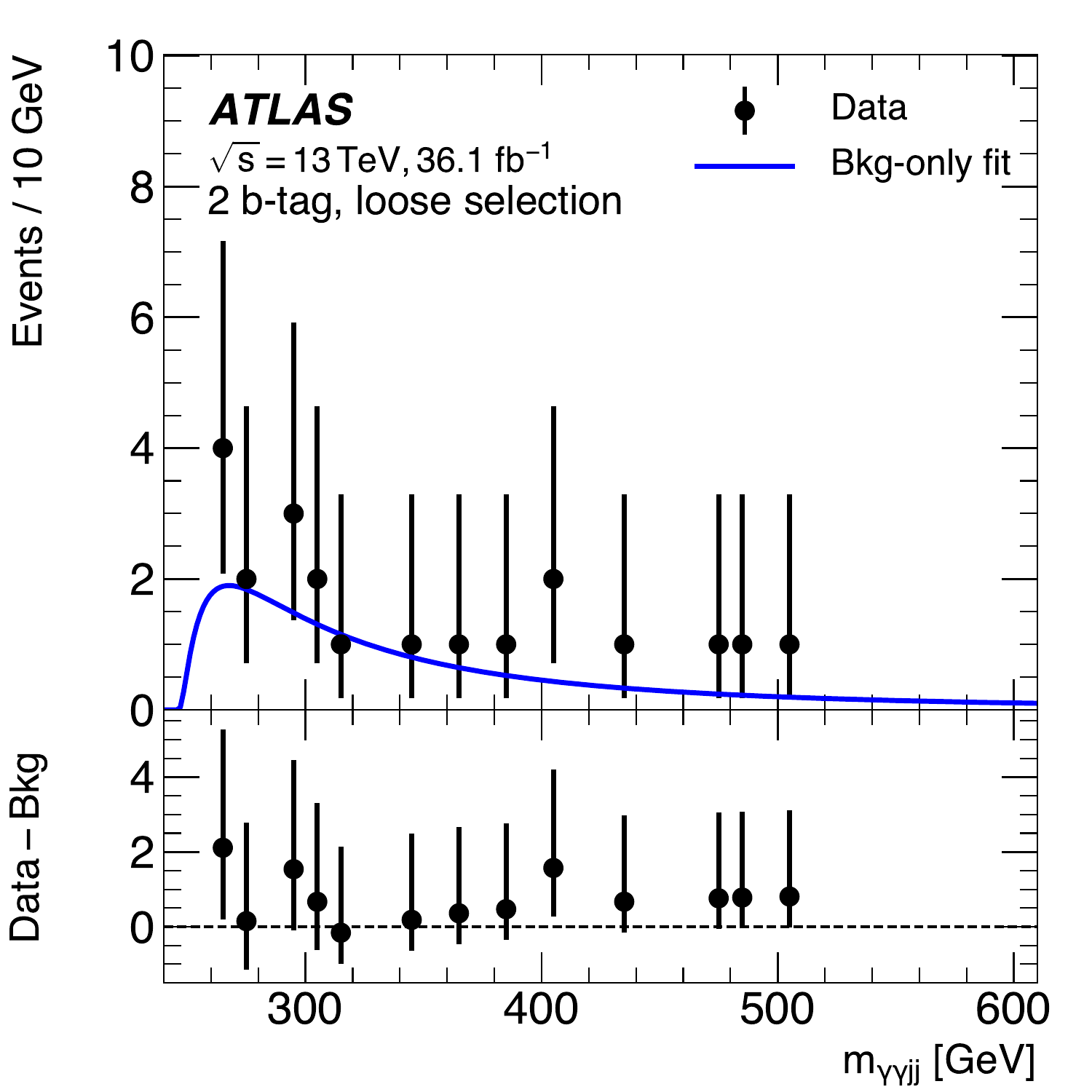}}
    \\
    \subfloat[]{\includegraphics[width=0.48\textwidth]{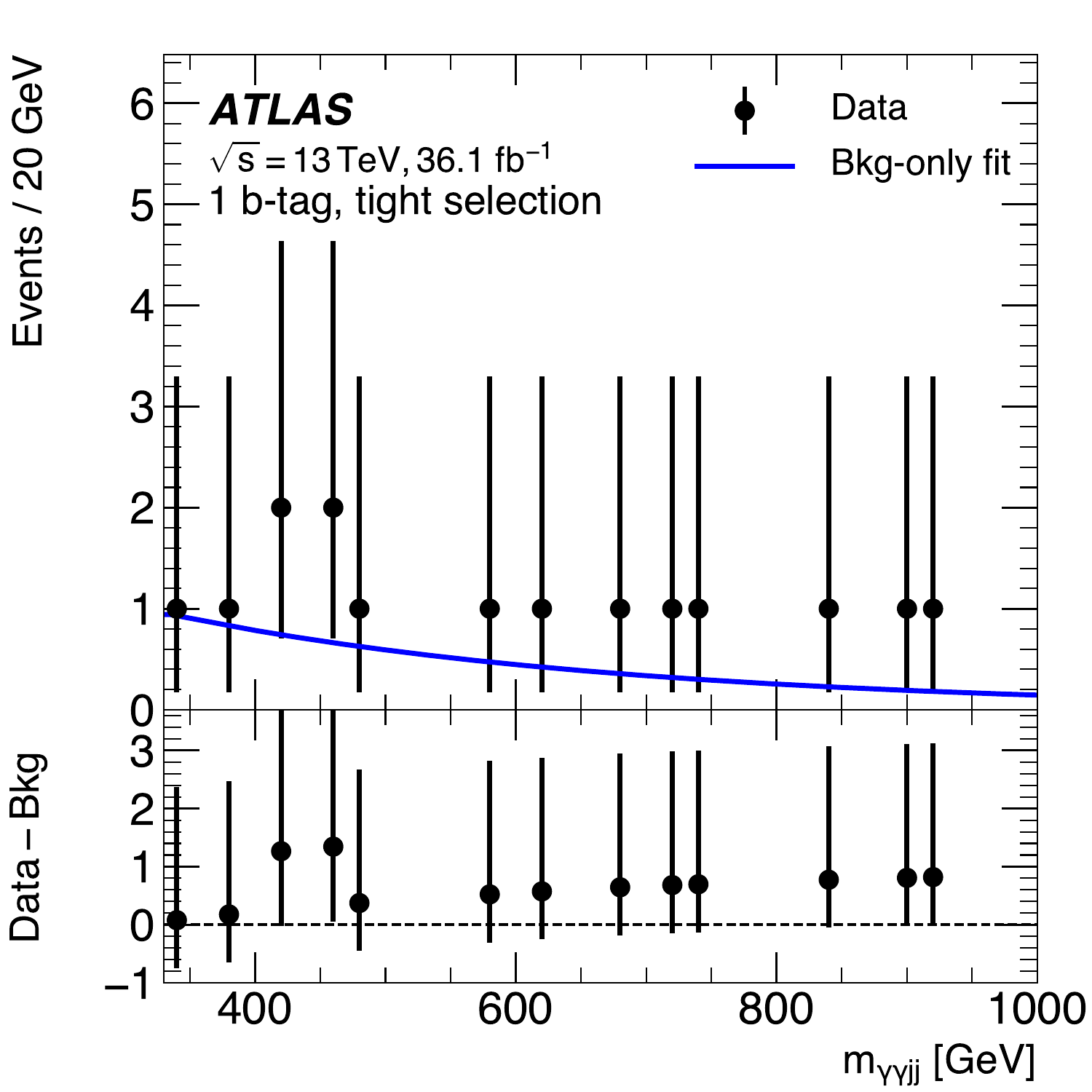}}
    \quad
    \subfloat[]{\includegraphics[width=0.48\textwidth]{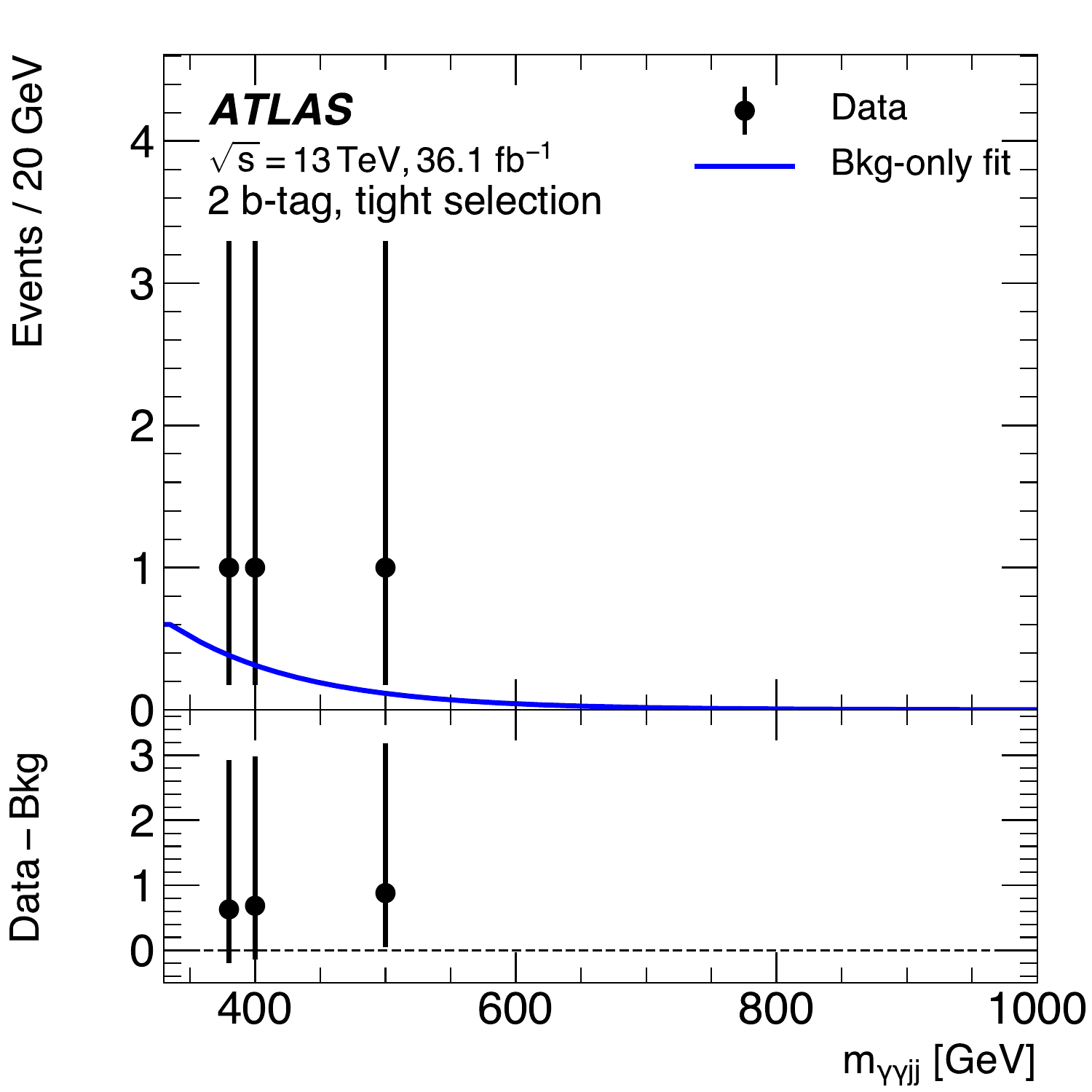}}
    \caption{For the resonant analysis, data (black points) are compared with the background-only fit (blue solid line) for \myyjj in the 1-tag (left) and 2-tag (right) categories with the loose (top) and tight (bottom) selections.
        The lower panel shows the residuals between the data and the best-fit background.}
    \label{results-myyjj}
\end{figure}

\Fig{\ref{results-myy}} shows the observed diphoton invariant mass spectra for the non-resonant analysis with the loose (top) and tight (bottom) selections.
The best-fit Higgs boson pair \xs is \SI[parse-numbers = false]{0.04_{-0.36}^{+0.43}~(-0.21_{-0.25}^{+0.33})}{\pico\barn} for the loose (tight) selection.
\Fig{\ref{results-myyjj}} shows the observed four-body invariant mass spectra for the resonant analysis in the loose (top) and tight (bottom) selections.
Maximum-likelihood background-only fits are also shown.
While local fluctuations may appear in some of the categories shown in \Fig{\ref{results-myyjj}}, only the combined two-category unbinned likelihood is considered for setting limits.
The largest discrepancy between the background-only hypothesis and the data manifests as an excess at \SI{480}{\GeV} with a local significance of $1.2~\sigma$.
The results are also interpreted as upper limits on the relevant Higgs boson pair production \xs{s}.

Exclusion limits are set on Higgs boson pair production in the \yybb final state.
The limits for both resonant and non-resonant production are calculated using the \CLs method~\cite{Read:2002hq}, with the likelihood-based test statistic $\tilde{q}_\mu$ which is suitable when considering signal strength $\mu\geq0$~\cite{Cowan:2010js,ATL-PHYS-PUB-2011-011}.
Because both the expected and observed numbers of events are small in the case of the resonant analysis, test-statistic distributions are evaluated by pseudo-experiments generated by profiling the nuisance parameters of the likelihood model on the observed data, as described in \Ref{\cite{ATL-PHYS-PUB-2011-011}}.
Better limits on \khhh are expected with the loose selection, whereas for the SM value $\khhh = 1$ the strongest limits on the Higgs boson pair \xs are derived from the tight selection.

\subsection{Exclusion limits on non-resonant \hh production}\label{sec:results_nonresonant_exclusion}
The 95\% confidence level (CL) upper limit for the non-resonant Higgs boson pair \xs is obtained using the tight selection.
\Fig{\ref{fig:results-non-resonant-limits}(a)} shows this upper limit, together with $\pm1\sigma$ and $\pm2\sigma$ uncertainty bands.
The observed (expected) value is \SI[parse-numbers = false]{\nonresobs~(\nonresexp)}{\pico\barn}.
As a multiple of the SM production \xs, the observed (expected) limits are \nonresnSigmaobs~(\nonresnSigmaexp).
The limits and the $\pm1\sigma$ band around each expected limit are presented in \Tab{\ref{tab:results-limitsWrtSM}}.

\begin{table}[!htb]
    \begin{center}
    \caption{The 95\% CL observed and expected limits on the Higgs boson pair \xs in \si{\pico\barn} and as a multiple of the SM production \xs.
             The $\pm1\sigma$ band around each 95\% CL limit is also indicated.}
    \label{tab:results-limitsWrtSM}
    \sisetup{separate-uncertainty, table-format = 1.3(4), table-align-exponent = false}
        \begin{tabular}{l c c c c}
            \toprule
                                                  & \multicolumn{1}{c}{Observed} & \multicolumn{1}{c}{Expected} & \multicolumn{1}{c}{$-1 \sigma$} & \multicolumn{1}{c}{$+1 \sigma$} \\
            \midrule
            \xsHH [\si{\pico\barn}]               & \nonresobs                   & \nonresexp                   & \nonresexpDown                  & \nonresexpUp                    \\
            & & \\ [-0.75em]
            As a multiple of $\sigma_\mathrm{SM}$ & \nonresnSigmaobs             & \nonresnSigmaexp             & \nonresnSigmaexpDown            & \nonresnSigmaexpUp              \\
            \bottomrule
        \end{tabular}
    \end{center}
\end{table}

\subsection{Exclusion limits on \texorpdfstring{\lhhh}{lambdahhh}}
\begin{figure}[!htb]
    \centering
    \subfloat[]{\includegraphics[width=0.48\textwidth]{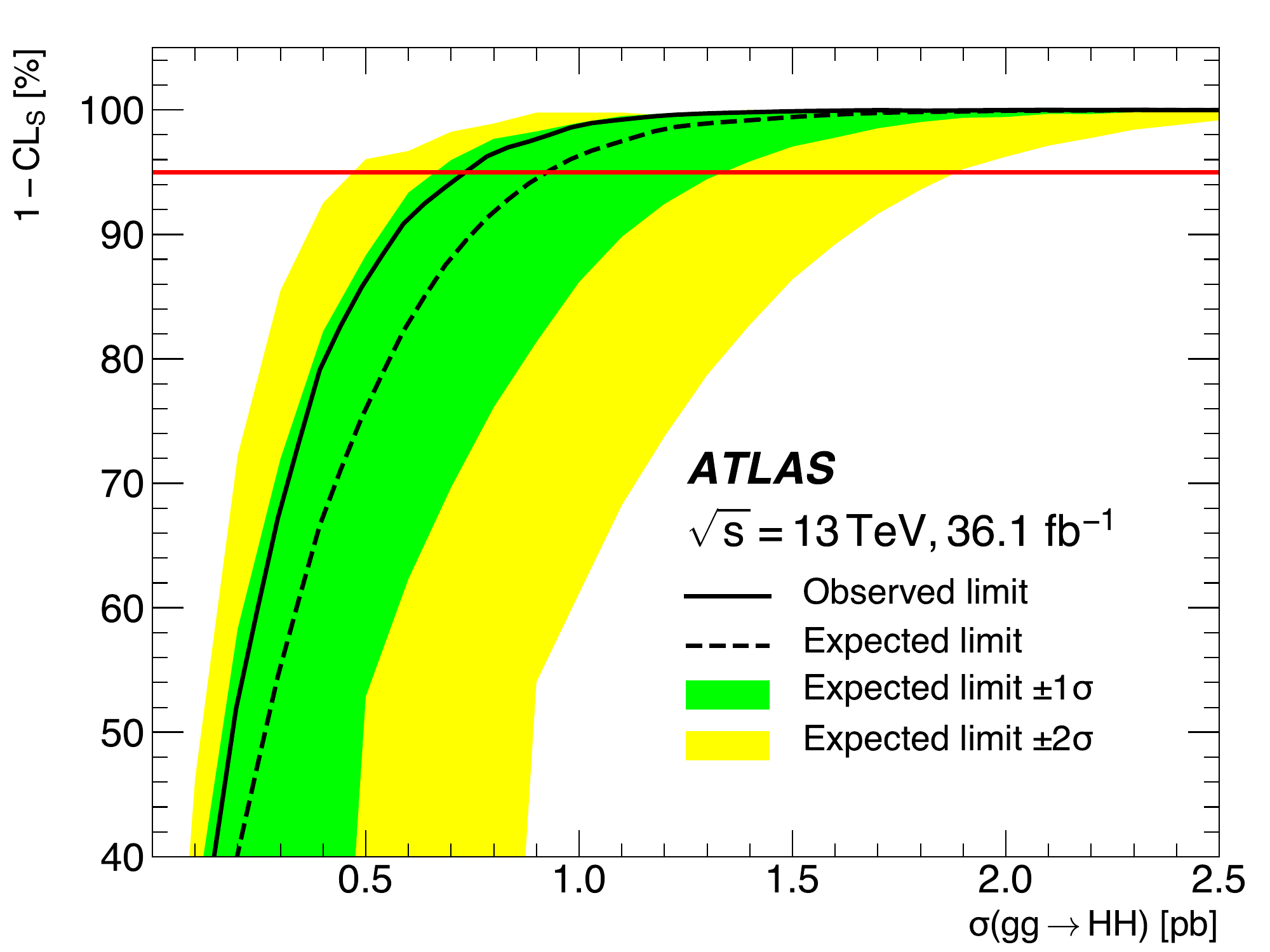}}
    \quad
    \includegraphics[width=0.48\textwidth]{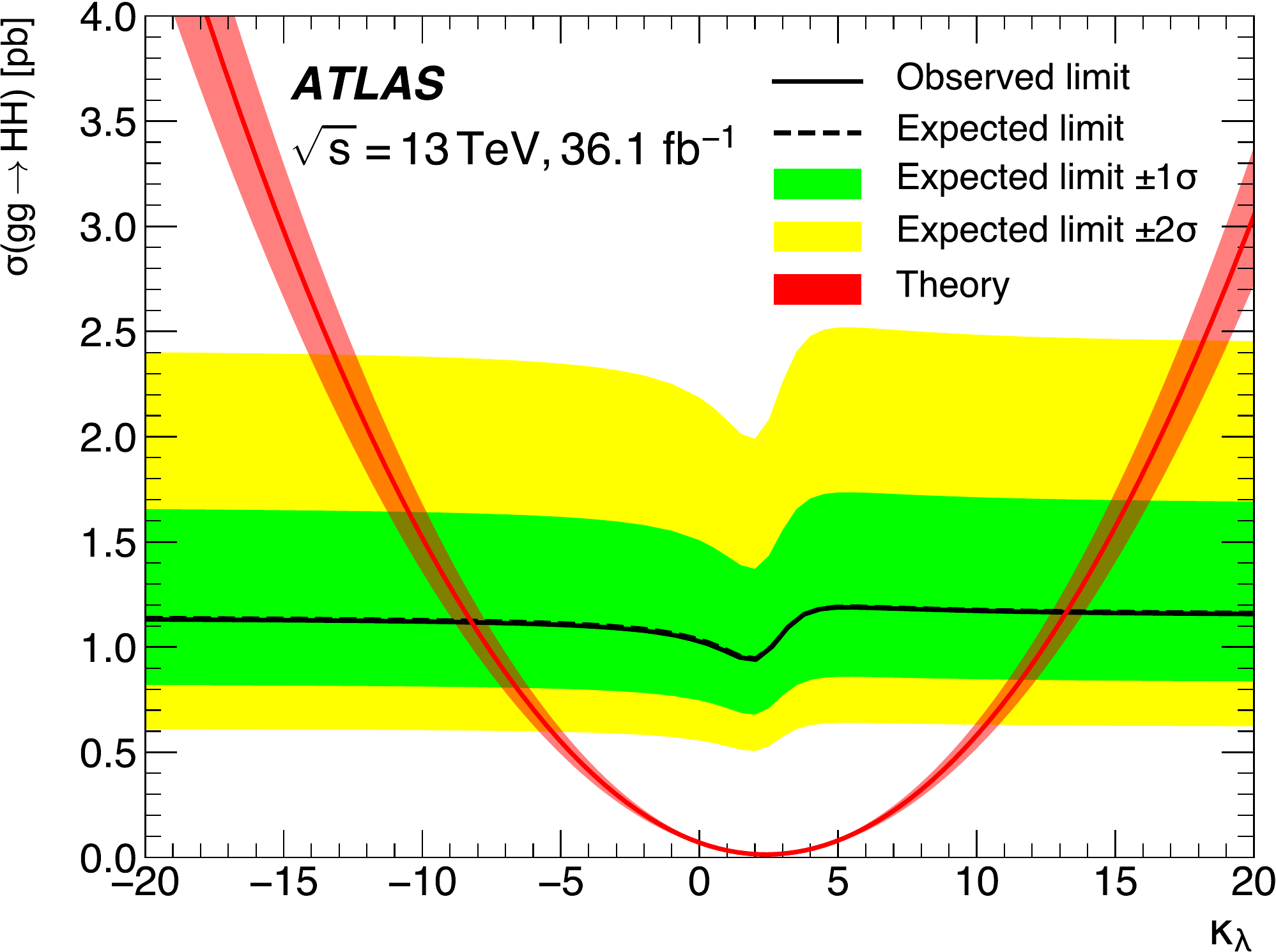}
    \caption{The expected and observed 95\% CL limits on the non-resonant production \xs \xsHH (a) for the SM-optimised limit using the tight selection and (b) as a function of $\kappa_\lambda$ using the loose selection.
        In (a) the red line indicates the 95\% confidence level.
        The intersection of this line with the observed, expected, and $\pm1\sigma$ and $\pm2\sigma$ bands is the location of the limits.
        In (b) the red line indicates the predicted \hh \xs if \khhh is varied but all other couplings remain at their SM values.
        The red band indicates the theoretical uncertainty of this prediction.}
    \label{fig:results-non-resonant-limits}
\end{figure}

Varying the Higgs boson self-coupling, \lhhh, affects both the total \xs of the non-resonant Higgs boson pair production and the event kinematics, affecting the signal selection efficiency.
In the non-resonant analysis, results are interpreted in the context of \khhh, using the loose selection, which is more sensitive for the range of \khhh values accessible with this data set.
As discussed in \Sect{\ref{subsec:data_MC_samples}}, the samples used for this interpretation were generated at \LO.
The 95\% CL limits on \xsHH are shown together with $\pm1\sigma$ and $\pm2\sigma$ uncertainty bands around the expected limit in \Fig{\ref{fig:results-non-resonant-limits}(b)}.
The limits are calculated using the asymptotic approximation~\cite{Cowan:2010js} for the profile-likelihood test statistic.
Fixing all other SM parameters to their expected values, the Higgs boson self-coupling is constrained at 95\% CL to $\lambdaminobs < \khhh < \lambdamaxobs$ whereas the expected limits are $\lambdaminexp <  \khhh < \lambdamaxexp$.

\subsection{Exclusion limits on resonant \hh production}
\label{sec:results_resonance_exclusion}
\begin{figure}[!htb]
    \centering
    \includegraphics[width=\textwidth]{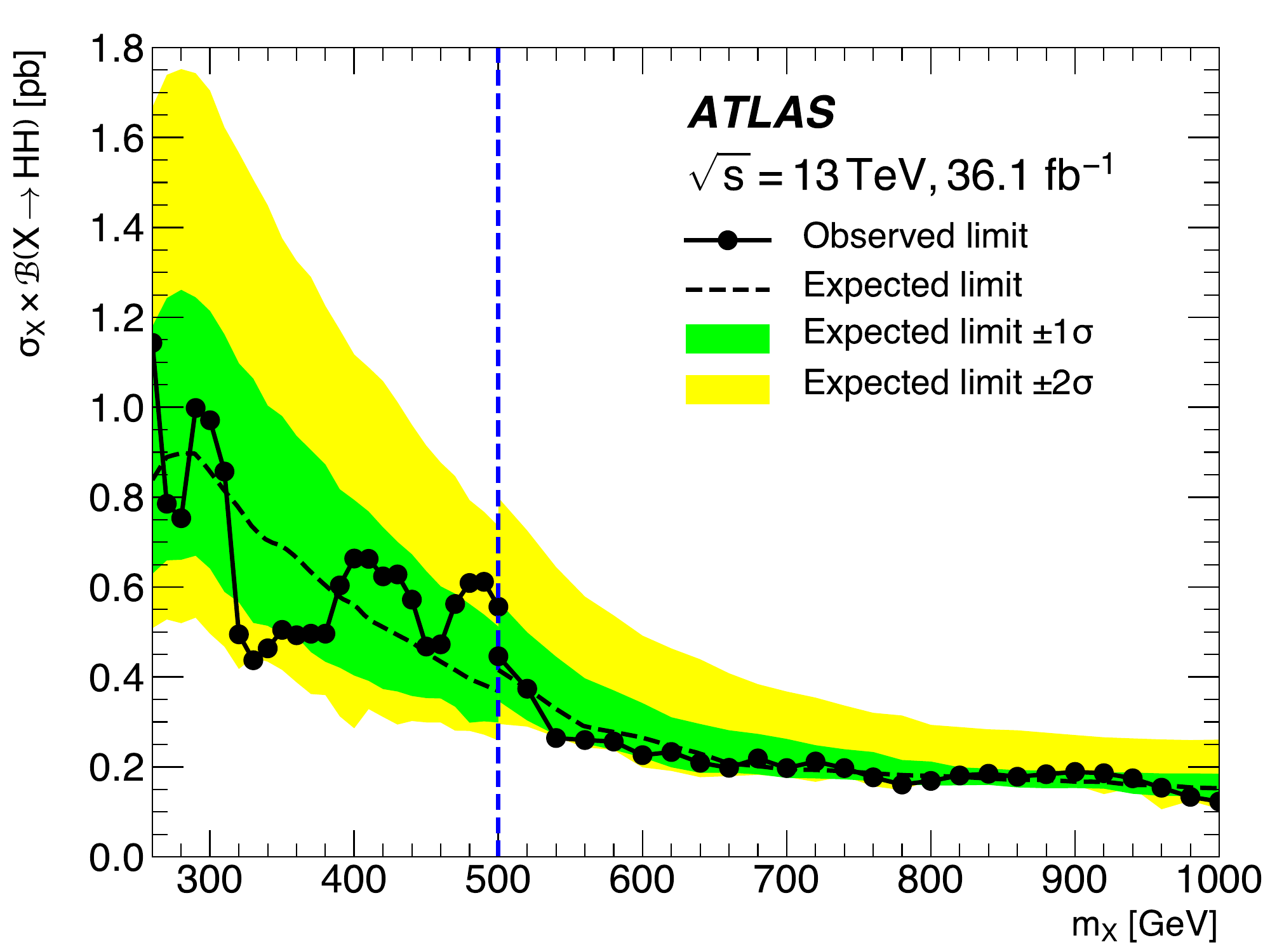}
    \caption{The expected and observed 95\% CL limits on the resonant production \xs, $\xsX \times \mathcal{B}\left(X \rightarrow \hh\right)$ as a function of \mX.
        The loose selection is used for $\mX \leq 500\GeV$, while the tight selection is used for $\mX \geq 500\GeV$.
        This is delineated with the blue dashed line.}
    \label{fig:results-resonant-limits}
\end{figure}

The 95\% CL limits on resonant Higgs boson pair production are shown in \Fig{\ref{fig:results-resonant-limits}}, utilising both the loose and tight selections.
The SM \hh contribution is considered as part of the background in this search although its inclusion has a negligible impact on the results.
For resonance masses in the range $\SI{260}{\GeV} < \mX < \SI{1000}{\GeV}$, the observed (expected) limits range between \SI[parse-numbers=false]{1.14~(0.90)}{\pico\barn} and \SI[parse-numbers=false]{0.12~(0.15)}{\pico\barn}.

\FloatBarrier

\section{Conclusions}\label{sec:conclusions}
Searches for resonant and non-resonant Higgs boson pair production in the \yybb final state are performed using \lumiInFemtoBarn of \pp collision data collected at $\sqs = \SI{13}{\TeV}$ with the ATLAS detector at the LHC in 2015 and 2016.
No significant deviations from the Standard Model predictions are observed.
A 95\% CL upper limit of \SI{\nonresobs}{\pico\barn} is set on the \xs for non-resonant production, while the expected limit is \SI{\nonresexp}{\pico\barn}.
This observed (expected) limit is \nonresnSigmaobs~(\nonresnSigmaexp) times the predicted SM \xs.
The Higgs boson self-coupling is constrained at 95\% CL to $\lambdaminobs < \khhh < \lambdamaxobs$ whereas the expected limits are $\lambdaminexp < \khhh < \lambdamaxexp$.
For resonant production of \Xhhyybb, a limit is presented for the narrow-width approximation as a function of \mX.
The observed (expected) limits range between \SI{\resmaxobs}{\pico\barn}~(\SI{\resmaxexp}{\pico\barn}) and \SI{\resminobs}{\pico\barn}~(\SI{\resminexp}{\pico\barn}) in the range $\SI{260}{\GeV} < \mX < \SI{1000}{\GeV}$.


\section*{Acknowledgements}

We thank CERN for the very successful operation of the LHC, as well as the
support staff from our institutions without whom ATLAS could not be
operated efficiently.

We acknowledge the support of ANPCyT, Argentina; YerPhI, Armenia; ARC, Australia; BMWFW and FWF, Austria; ANAS, Azerbaijan; SSTC, Belarus; CNPq and FAPESP, Brazil; NSERC, NRC and CFI, Canada; CERN; CONICYT, Chile; CAS, MOST and NSFC, China; COLCIENCIAS, Colombia; MSMT CR, MPO CR and VSC CR, Czech Republic; DNRF and DNSRC, Denmark; IN2P3-CNRS, CEA-DRF/IRFU, France; SRNSFG, Georgia; BMBF, HGF, and MPG, Germany; GSRT, Greece; RGC, Hong Kong SAR, China; ISF and Benoziyo Center, Israel; INFN, Italy; MEXT and JSPS, Japan; CNRST, Morocco; NWO, Netherlands; RCN, Norway; MNiSW and NCN, Poland; FCT, Portugal; MNE/IFA, Romania; MES of Russia and NRC KI, Russian Federation; JINR; MESTD, Serbia; MSSR, Slovakia; ARRS and MIZ\v{S}, Slovenia; DST/NRF, South Africa; MINECO, Spain; SRC and Wallenberg Foundation, Sweden; SERI, SNSF and Cantons of Bern and Geneva, Switzerland; MOST, Taiwan; TAEK, Turkey; STFC, United Kingdom; DOE and NSF, United States of America. In addition, individual groups and members have received support from BCKDF, the Canada Council, CANARIE, CRC, Compute Canada, FQRNT, and the Ontario Innovation Trust, Canada; EPLANET, ERC, ERDF, FP7, Horizon 2020 and Marie Sk{\l}odowska-Curie Actions, European Union; Investissements d'Avenir Labex and Idex, ANR, R{\'e}gion Auvergne and Fondation Partager le Savoir, France; DFG and AvH Foundation, Germany; Herakleitos, Thales and Aristeia programmes co-financed by EU-ESF and the Greek NSRF; BSF, GIF and Minerva, Israel; BRF, Norway; CERCA Programme Generalitat de Catalunya, Generalitat Valenciana, Spain; the Royal Society and Leverhulme Trust, United Kingdom.

The crucial computing support from all WLCG partners is acknowledged gratefully, in particular from CERN, the ATLAS Tier-1 facilities at TRIUMF (Canada), NDGF (Denmark, Norway, Sweden), CC-IN2P3 (France), KIT/GridKA (Germany), INFN-CNAF (Italy), NL-T1 (Netherlands), PIC (Spain), ASGC (Taiwan), RAL (UK) and BNL (USA), the Tier-2 facilities worldwide and large non-WLCG resource providers. Major contributors of computing resources are listed in Ref.~\cite{ATL-GEN-PUB-2016-002}.


\printbibliography

\clearpage
 
\begin{flushleft}
{\Large The ATLAS Collaboration}

\bigskip

M.~Aaboud$^\textrm{\scriptsize 34d}$,    
G.~Aad$^\textrm{\scriptsize 99}$,    
B.~Abbott$^\textrm{\scriptsize 125}$,    
O.~Abdinov$^\textrm{\scriptsize 13,*}$,    
B.~Abeloos$^\textrm{\scriptsize 129}$,    
D.K.~Abhayasinghe$^\textrm{\scriptsize 91}$,    
S.H.~Abidi$^\textrm{\scriptsize 164}$,    
O.S.~AbouZeid$^\textrm{\scriptsize 39}$,    
N.L.~Abraham$^\textrm{\scriptsize 153}$,    
H.~Abramowicz$^\textrm{\scriptsize 158}$
% ...
\bigskip
\\

% ...
$^{13}$Institute of Physics, Azerbaijan Academy of Sciences, Baku; Azerbaijan.\\
% ...
$^{34}$$^{(a)}$Facult\'e des Sciences Ain Chock, R\'eseau Universitaire de Physique des Hautes Energies - Universit\'e Hassan II, Casablanca;$^{(b)}$Centre National de l'Energie des Sciences Techniques Nucleaires (CNESTEN), Rabat;$^{(c)}$Facult\'e des Sciences Semlalia, Universit\'e Cadi Ayyad, LPHEA-Marrakech;$^{(d)}$Facult\'e des Sciences, Universit\'e Mohamed Premier and LPTPM, Oujda;$^{(e)}$Facult\'e des sciences, Universit\'e Mohammed V, Rabat; Morocco.\\
% ...
$^{39}$Niels Bohr Institute, University of Copenhagen, Copenhagen; Denmark.\\
% ...
$^{91}$Department of Physics, Royal Holloway University of London, Egham; United Kingdom.\\
% ...
$^{99}$CPPM, Aix-Marseille Universit\'e, CNRS/IN2P3, Marseille; France.\\
% ...
$^{125}$Homer L. Dodge Department of Physics and Astronomy, University of Oklahoma, Norman OK; United States of America.\\
% ...
$^{129}$LAL, Universit\'e Paris-Sud, CNRS/IN2P3, Universit\'e Paris-Saclay, Orsay; France.\\
% ...
$^{153}$Department of Physics and Astronomy, University of Sussex, Brighton; United Kingdom.\\
% ...
$^{158}$Raymond and Beverly Sackler School of Physics and Astronomy, Tel Aviv University, Tel Aviv; Israel.\\
% ...
$^{164}$Department of Physics, University of Toronto, Toronto ON; Canada.\\
% ...
$^{a}$ Also at Borough of Manhattan Community College, City University of New York, NY; United States of America.\\
$^{b}$ Also at California State University, East Bay; United States of America.\\
$^{c}$ Also at Centre for High Performance Computing, CSIR Campus, Rosebank, Cape Town; South Africa.\\
$^{d}$ Also at CERN, Geneva; Switzerland.\\
$^{e}$ Also at CPPM, Aix-Marseille Universit\'e, CNRS/IN2P3, Marseille; France.\\
% ...
$^{*}$ Deceased

\end{flushleft}



%

\end{document}